\documentclass[a4paper,12pt]{article}

\usepackage{amsmath}
\usepackage{amssymb}
\usepackage[dvips]{graphicx}
\usepackage[dvips]{psfrag}

\newcommand{\be}{\begin{equation}}
\newcommand{\ee}{\end{equation}}
\newcommand{\bea}{\begin{eqnarray}}
\newcommand{\eea}{\end{eqnarray}}
\newcommand{\nono}{\nonumber}
\newcommand{\tr}{{\rm tr}}

\newcommand{\vev}[1]{\left\langle #1 \right\rangle}

\newcommand{\Slash}[1]{{\ooalign{\hfil/\hfil\crcr$#1$}}}

\begin{document}
\begin{titlepage}
\begin{flushright}
\begin{tabular}{l}
KEK-TH-1195\\
OIQP-07-14\\
\end{tabular}
\end{flushright}

\vspace{5mm}

\begin{center}
{\Large \bf Spontaneous Supersymmetry Breaking\\
 by Large-$N$ Matrices}
\baselineskip=24pt

\vspace{15mm}
\large
Tsunehide Kuroki,\footnote{tkuroki@post.kek.jp}\\
{\it High Energy Accelerator Research Organization (KEK), \\
Tsukuba, Ibaraki 305-0801, Japan}\\
and\\
Fumihiko Sugino,\footnote{fumihiko\_sugino@pref.okayama.lg.jp}\\
{\it Okayama Institute for Quantum Physics, \\
Kyoyama 1-9-1, Okayama 700-0015, Japan}
\vspace{20mm}
\end{center}
\begin{abstract}
Motivated by supersymmetry breaking in matrix model formulations of superstrings, 
we present some concrete models, in which the supersymmetry is preserved for any finite $N$, 
but gets broken at infinite $N$, where $N$ is the rank of matrix variables. 
The models are defined as supersymmetric field theories coupled to some matrix models, 
and in the induced action obtained after integrating out the matrices, 
supersymmetry is spontaneously broken 
only when $N$ is infinity. 
In our models, the large value of $N$ gives a natural explanation 
for the origin of small parameters 
appearing in the field theories which trigger the supersymmetry breaking.  
 
In particular, in the case of the O'Raifeartaigh model coupled to 
a certain supersymmetric matrix model, 
a nonsupersymmetric metastable vacuum appears near the origin of the field space, which is far from 
the position of the supersymmetric vacuum.   
We estimate its lifetime as a function of $N$. 

\end{abstract}
\end{titlepage}

\section{Introduction}
\label{sec:intro}
Matrix model formulations have many intriguing features suitable for fully nonperturbative description of superstring theory. 
As one of such aspects, matrices originally arise as regularized one-string (one-membrane) coordinates, 
but remarkably they can contain multi-string (multi-membrane) 
states as sub-blocks in the matrices in a quite natural manner~\cite{Ishibashi:1996xs,Banks:1996vh}. 
In the limit that the size of the matrices $N$ becomes infinity, the matrix models are expected to give 
a constructive definition of the superstring theory. 
However, the correct large-$N$ limit has not been made clear yet\footnote{The recent paper~\cite{Kawai:2007tn} 
gives an interesting analysis of the large-$N$ limit.}. 
  
As for the IIB matrix model~\cite{Ishibashi:1996xs}, 
Gaussian expansion or improved mean-field analysis strongly suggests 
that the four-dimensional macroscopic space-time emerges 
as a consequence of spontaneous breaking of ten-dimensional Lorentz 
symmetry~\cite{Nishimura:2001sx}.  
Also, it is desirable to reveal geometric structure of the resulting space-time 
and to explain the emergence of matter fields and their internal (gauge) symmetries. 
If the matrix model formulation of superstring theory like 
the IIB matrix model is true and describes the real world, it must reproduce the standard model 
in the low energy regime. From this point of view, it is expected 
that symmetries of the original matrix model such as U($N$) symmetry 
and supersymmetries should be (partly) broken in the large-$N$ limit. 
Moreover, spontaneous symmetry breaking in the large-$N$ limit 
would be quite interesting in itself from the theoretical point of view.   

As one of these topics, in this paper, we focus on spontaneous supersymmetry (SUSY) breaking in the context of matrix models. 
The IIB matrix model with finite $N$, which is defined by finite-dimensional matrix integrals, does not have any source of  
SUSY breaking. 
Both of the classical action and the path-integral measure is supersymmetric, 
and the finite-dimensional integrals are well-defined~\cite{Austing:2001pk} with no subtlety concerning regularizations.
Thus, 
we will not observe any irregular behavior in the finite-dimensional integrals, 
which can violate supersymmetric Ward-Takahashi identities. 
Furthermore, there arises no superselection sector because of 
finiteness of the system. 
Therefore, if SUSY is expected to be broken in the IIB matrix model, 
it must occur only in the large-$N$ limit. 
Namely, supersymmetry should be preserved for any finite $N$, 
but it is spontaneously broken at $N=\infty$. 
Hence, order parameters for the SUSY breaking do not behave smoothly 
at large $N$.    

Such situation is rather unusual, and to the best of our knowledge, 
it has not ever been discussed in the literature. 
In ref.~\cite{Witten:1982df}, Witten discussed the case 
that SUSY is spontaneously broken in any finite volume, but restored in the 
infinite volume limit, by using a ($1+1$)-dimensional simple model. 
Typically, SUSY breaking is triggered by instantons, and their contribution is estimated as ${\cal O}(e^{-{\cal V}})$ 
(${\cal V}$ is the spatial volume), 
which vanishes in the infinite volume limit. 
Also, various models for SUSY breaking in the large-$N$ limit 
were investigated~\cite{Zanon:1981td,Davis:1982ad,Affleck:1983pn,
Higashijima:1983er,Yahikozawa:1986pt}, 
however in these models the SUSY is considered to be already broken 
at finite $N$, 
because we can not find any specific mechanism 
which spoils the smoothness of the large-$N$ limit and 
makes discontinuity of the vacuum energy at $N=\infty$. 
Similarly to this case, Affleck~\cite{Affleck:1983pn} discussed 
a possibility 
that the SUSY breaking occurring for finite $N$ ceases 
in the large-$N$ limit as ${\cal O}(e^{-N})$.    

In this paper, we construct some concrete models realizing our desirable situation, 
in which SUSY is preserved at any finite $N$ and SUSY breaking takes place only at $N=\infty$. 
These models are of the form of supersymmetric field theories 
coupled to some zero-dimensional large-$N$ matrix models, 
and the total actions are also supersymmetric. 
Their couplings are introduced by promoting some parameters in the field theories to dynamical variables in the 
matrix models. 
They can be regarded as supersymmetric field theories 
with SUSY breaking hidden sectors composed of matrix variables, 
which become important only at large $N$. 
Strictly speaking, our models are not of the form of single matrix models 
since they contain field theory sectors, but we expect that 
they will be relevant for future investigations in the matrix model 
formulations of superstring theory in the following sense: 
we could consider packing the two sectors -- field theory (FT) sector and matrix model (MM) sector -- into 
a single matrix as 
\be
{\cal A} = \left(\begin{array}{cc}
\mbox{FT sector} & \mbox{interaction} \\
\mbox{interaction} & \mbox{MM sector}
\end{array}
\right), 
\label{big_matrix}
\ee
where the FT sector is properly discretized to fit the matrix description, and ``interaction" in the off-diagonal blocks gives the 
interaction terms between the two sectors. 
In the perspective of the IIB matrix model, it would be intriguing to discover such a single matrix model of ${\cal A}$ deriving the models 
appearing in following sections.    

In section~\ref{sec:proto}, we discuss a simple two-dimensional supersymmetric model coupled to the one-matrix model. 
We show that the SUSY breaking occurs only at $N=\infty$ 
for a quite general potential of the matrix model. 
In section~\ref{sec:ORai}, the second example realizing our desirable situation is presented. 
We take the O'Raifeartaigh model as the FT sector, to which a supersymmetric matrix model is coupled. 
We analyze the case where the potential of the matrix model 
is Gaussian, and 
the induced action obtained after integrating out the matrices exhibits SUSY breaking only 
at infinite $N$. 
In section~\ref{sec:meta}, the analysis in the previous section is extended to the case where the matrix model has cubic interactions. 
It also realizes the desirable SUSY breaking, 
and thus it is expected that the mechanism works in more general matrix model potentials.  
Next, by taking into account the one-loop effects of the induced action obtained in section~\ref{sec:ORai}, 
we see that a metastable vacuum appears near the origin of the field space.    
The lifetime of the vacuum is estimated as a function of $N$, and we find that it becomes longer as $N$ increases. 
In section~\ref{sec:outlook}, we summarize the results so far and explain their applications to other settings and 
to some naturalness problems. 
In appendix~\ref{app:proto}, a proof is given that 
the supersymmetry is preserved in finite $N$ case 
in the first example of section~\ref{sec:proto}. 
Appendix~\ref{app:loop_susymat} is devoted to computational details of the supersymmetric matrix model with 
cubic interactions. 
It is also discussed that the coupling between the FT sector and the MM sector is consistent in the Euclid space, 
but can be inconsistent in the Minkowski space.  
Finally, some details in computing the vacuum decay rate and remarks on the bounce calculation for nonlocal potentials 
are presented in appendix~\ref{app:lifetime}.

\section{Witten's Simple Model Coupled to One-Matrix Model}
\label{sec:proto}
\setcounter{equation}{0}
As the first example, we consider a simple two-dimensional supersymmetric theory as the FT sector: 
\be
S_{FT} =  \int d^2x\left[\frac12(\partial_{\mu}\phi)^2 + \frac12\bar{\psi}\Slash{\partial}\psi 
+\frac{\lambda^2}{2}(\phi^2+a^2)^2 + \lambda\phi\bar{\psi}\psi\right], 
\label{pro_FT}
\ee
and the one-matrix model as the MM sector:
\be
S_{MM}  = N \tr \, V(M).  
\label{MMpot}
\ee 
Here, $\phi$ is a real scalar field, $\psi$ is a two-component Majorana spinor, and the space-time is Wick-rotated:
\be
\{\gamma_{\mu}, \gamma_{\nu}\} = 2\delta_{\mu\nu}. 
\ee
Also, $M$ is an $N\times N$ hermitian matrix and we make a mild 
assumption that the potential $V(M)$ is even and bounded from below, 
and that $e^{-NV(x)}$ damps exponentially as $|x| \to \infty$. 
The action of the total system is given by 
\be
S_{tot} = S_{FT} + S_{MM}. 
\label{total_system}
\ee 
As a coupling between the two sectors, we take the constant $a^2$ in (\ref{pro_FT}) to be $M$-dependent as 
\be
a^2= \frac{m^2}{2\lambda^2}\left\{(1+\kappa )a_0^2 -\kappa\right\}, \qquad 
a_0^2 \equiv \frac{\epsilon^2}{(\frac{1}{N}\tr M)^2 + \epsilon^2}, 
\label{pro_coupling}
\ee     
where $\kappa$ is fixed to an arbitrary positive constant, 
and $\epsilon$ is a real parameter finally sent to zero. 
$S_{tot}$ is invariant under the SUSY transformation with the parameter $\xi$: 
\bea
\delta_{\xi} \phi & = & \bar{\xi}\psi, \nono \\
\delta_{\xi} \psi & = & -(\Slash{\partial}\phi + h'(\phi))\xi, \qquad 
h'(\phi) = \lambda (\phi^2+a^2), \nono \\
\delta_{\xi}M & = & 0. 
\label{pro_SUSYtr}
\eea

$S_{FT}$ is an Euclidean version of the action discussed in ref.~\cite{Witten:1982df}, where SUSY is broken both for $a^2>0$ and  
for $a^2< 0$ in any finite volume (length) of spatial direction, but the SUSY is restored for $a^2< 0$ in the infinite volume. 
For the case $a^2 < 0$, $\phi$ has two minima $\phi = \pm\sqrt{-a^2}$. The instanton effect 
causing tunneling between them triggers the supersymmetry breaking 
in finite spatial volume. 
Since the amount of the effect is ${\cal O} (e^{-{\cal V}})$ (${\cal V}$ is the spatial volume), 
it vanishes in the infinite volume and the SUSY is restored. 

Here, we integrate out the MM-sector variable $M$ and show that the result causes the SUSY breaking only at $N=\infty$.  
In the matrix integral with respect to the action $S_{MM}$, 
\be
\vev{\left(\frac{1}{N}\tr \, M\right)^2} = \left(\vev{\frac{1}{N}\tr \, M}\right)^2 + \vev{\left(\frac{1}{N}\tr \, M\right)^2}_C, 
\ee
where the symbol $\vev{\cdot}$ is used for the expectation value in the MM sector, 
and the suffix $C$ means taking contributions from the connected diagrams. 
{}From the $Z_2$-symmetry of $V(M)$ under $M \to -M$, 
the first term vanishes, and the second term remains for finite $N$, which also vanishes 
as ${\cal O}(1/N^2)$ in the large-$N$ limit. 
Likewise, in 
\be
\vev{a_0^2} = 1+ \sum_{k=1}^\infty \frac{1}{\epsilon^{2k}}\vev{\left(\frac{1}{N}\tr \, M\right)^{2k}}, 
\ee
each term in the summation vanishes in the large-$N$ limit. 
On the other hand, 
for finite $N$, we show in appendix~\ref{app:proto} that $\vev{a_0^2}\to 0$ as $\epsilon \to 0$. 
Thus, after taking $\epsilon\to 0$, we have 
\be
\vev{a_0^2} = \left\{ \begin{array}{cl} 0 & \mbox{for $N$: finite} \\
                                1 & \mbox{for $N=\infty$,} \end{array} \right. 
\label{pro_veva0}                                
\ee  
hence\footnote{It seems delicate to discuss the SUSY of $S_{FT}$ in the case $a^2=0$. 
To avoid such subtlety and for safety, we introduced the positive constant $\kappa$.} 
\be
\vev{a^2} = \left\{ \begin{array}{ll} -\frac{m^2}{2\lambda^2}\kappa < 0 & \mbox{for $N$: finite} \\
                                + \frac{m^2}{2\lambda^2} > 0 & \mbox{for $N=\infty$.} \end{array} \right. 
\label{pro_veva}                                
\ee  
For finite $N$, SUSY is broken by the amount of ${\cal O}(e^{-{\cal V}})$, but it is restored in the infinite volume 
limit ${\cal V}\to \infty$. 
On the other hand, at $N=\infty$, SUSY is broken for arbitrary ${\cal V}$.  
In the latter case, around the minimum $\phi=0$,  
$\phi$ has mass $m$, while $\psi$ is massless; $\psi$ is a Nambu-Goldstone fermion associated to the SUSY breaking.  

In order to make this argument complete, we have to take account 
of the higher-order contributions of 
$\vev{a^{2k}}$ ($k\ge 2$), which arise from expanding the exponential of 
$\exp\left[-\int d^2x\, \frac{\lambda^2}{2} (\phi(x)^2+ a^2)^2\right]$. 
However, it is shown in appendix~\ref{app:proto} that $\vev{a_0^{2k}} \to 0$ for finite $N$, and in the case of $N=\infty$ 
\be
\vev{a_0^{2k}}=\left(\vev{a_0^2}\right)^k 
\ee
holds from the large-$N$ factorization. 
Therefore, we can see that 
\be
\vev{\exp\left[-\int d^2x \, \frac{\lambda^2}{2}(\phi^2+a^2)^2\right]}=
\exp\left[-\int d^2x \, \frac{\lambda^2}{2}\left(\phi^2+\vev{a^2}\right)^2\right] 
\ee
irrespective of $N$ being finite or infinite, 
and that the above conclusion dose not change even after taking into account the higher-order effects.  

In this model, the jump of the value $\vev{a^2}$ at $N=\infty$ in (\ref{pro_veva}) is crucial for the large-$N$ 
SUSY breaking. It originates from changing the order of the limits $\epsilon\to 0$ and $N \to \infty$. 
In this sense, $\epsilon$ plays a similar role to an external source introduced in a system 
which exhibits spontaneous breaking of internal symmetry -- for example, 
magnetic field in the Ising model. 
It is well known that 
the thermodynamical limit and the limit of turning off the external 
source do not commute below the critical temperature\footnote{
However, in contrast to the fact that the external source explicitly 
breaks the symmetry, $\epsilon$ in our model does not break 
the SUSY explicitly. 
The point is that the SUSY transformation (\ref{pro_SUSYtr}) depends on $\epsilon$ 
through the expression of $a^2$ (\ref{pro_coupling}).  
We should also note this difference.}. 
Moreover, it is equally crucial that the higher-order effects of $\vev{a^{2k}}$ ($k\ge 2$) do not spoil the SUSY which is preserved 
for finite $N$ at the level of $\vev{a^2}$. 
We should remark that the limit $\epsilon\to 0$ is also important 
at this point.

Here we explain that it is quite nontrivial to maintain SUSY for finite $N$. 
In fact, instead of the MM-sector considered here, 
if we consider any system of size $N$, in which its internal symmetry 
is spontaneously broken, and simply 
take an order parameter of this symmetry breaking as $a^2$ in (\ref{pro_FT}), 
the desirable large-$N$ SUSY breaking might seem to be easily realized. 
In this case, since the internal symmetry breaking occurs at $N=\infty$ but not for any finite $N$, 
the jump of $\vev{a^2}$ at $N=\infty$ can be produced 
without difficulty. 
On the other hand, the higher-order correlators $\vev{a^{2k}}$ ($k\ge 2$) generically 
give nonvanishing results in {\em finite} $N$ 
and break the SUSY which seems retained at the first order. 
(For example, we may consider the two-dimensional Ising model with $N$-sites coupled to the FT-sector (\ref{pro_FT}) 
and take 
\be
a^2 = \frac{m^2}{2\lambda^2}\left\{ (1+\kappa) a_0^2 -\kappa\right\}, 
\qquad a_0^2 \equiv \frac{1}{N}\sum_i s_i, 
\ee
where $s_i=\pm 1$ is a spin variable at the site $i$. 
The temperature is set to any value below the critical temperature. 
The jump of $\vev{a^2}$ at $N=\infty$ is no problem, but the $K$-point correlators ($K\ge 2$) among spin variables 
give nonvanishing effects and easily violate the SUSY for {\em finite} $N$). 
Therefore, we can say that, in constructing that type of desirable models, 
it needs more consideration to maintain SUSY for finite $N$ than to 
realize the SUSY breaking at $N=\infty$. 

It is possible to make such SUSY breaking only in the FT sector 
without introducing the MM sector. 
For example, if we take
\be
a^2 = \frac{m^2}{2\lambda^2}\left\{ (1+\kappa) a_0^2 -\kappa\right\}, 
\qquad a_0^2\equiv  \frac{(N\epsilon)^2}{1 + (N\epsilon)^2}
\ee 
in $S_{FT}$ instead of (\ref{pro_coupling}), 
the model in the infinite volume 
preserves the SUSY for finite $N$, and exhibits the breaking at $N=\infty$. 
In this case, however, the parameter $N$ does not have a natural meaning in the FT sector alone. 
{}From the aspect of the naturalness, it will be better to introduce $N$ through coupling to 
some model with the degrees of freedom associated with $N$, for example matrix models or vector models. 
Note that the coupling (\ref{pro_coupling}) respects the U($N$) symmetry of the matrix model, 
whereas in the O($N$) or U($N$) vector models it is difficult to consider the counterpart of the ``parity-odd'' object $\tr M$ 
respecting their O($N$) or U($N$) symmetry. 
Thus, we can say that this kind of the breaking mechanism fits most naturally for the coupling to the matrix models, 
not to the vector models.  
It should be also noticed that the integral over $M$ plays 
an essential role in the discontinuity in (\ref{pro_veva0}). 

Next, let us try to express in a more familiar form the coupling (\ref{pro_coupling}), 
where the matrix $M$ appears in the denominator. 
Introducing a scalar field $X(x)$, we rewrite the coupling term as 
\bea
\lefteqn{\exp\left[-\int d^2x \frac{m^2}{2}(1+\kappa) a_0^2\phi(x)^2\right]}  \nono \\ 
& =&  
C(a_0^2) \int {\cal D}X \exp\left[-\int d^2x \left(\frac{1}{m^2(1+\kappa) a_0^2}X(x)^2 + i\sqrt{2}\phi(x)X(x)\right)\right] \nono \\
& = & \tilde{C}(a_0^2) \int {\cal D}X \exp\left[-\int d^2x 
\left\{\frac{\Lambda^4}{m^2(1+\kappa)}\left(1+\frac{1}{(N\epsilon)^2}(\tr M)^2\right)X(x)^2 \right.\right. \nono \\
 & & \hspace{5cm} \left.\left. + i\sqrt{2}\Lambda^2\phi(x)X(x)\right\}\right] 
\label{scaling} \\
 & = & 
 \tilde{C}(a_0^2) \int {\cal D}X \exp\left[-\int d^2x 
\left(\frac12 \mu(M)^2 X(x)^2 + i\sqrt{2}\Lambda^2\phi(x)X(x)\right)\right] 
\label{rewrite1}
\eea
with 
\be
\mu(M)^2 \equiv \frac{2\Lambda^4}{m^2(1+\kappa)}\left(1+\frac{1}{(N\epsilon)^2}(\tr M)^2\right). 
\ee 
$C(a_0^2)$ and $\tilde{C}(a_0^2)$ are $a_0^2$-dependent constants, 
and $X(x)$ was rescaled as $X(x) \to \Lambda^2 X(x)$ in (\ref{scaling}) 
to have the canonical dimension. 
Also, we have to introduce a supersymmetric partner of $X(x)$, which is a two-component Majorana spinor (call $\Xi(x)$), 
to cancel the prefactor $\tilde{C}(a_0^2)$ in (\ref{rewrite1}). As a result, 
\bea
 & & \exp\left[-\int d^2x \frac{m^2}{2}(1+\kappa) a_0^2\phi(x)^2\right] =
\int {\cal D}X {\cal D}\Xi \, e^{-S_{mess}}, \\
 & & S_{mess}  \equiv  \int d^2 x \left[\frac12 \mu(M)^2X(x)^2 +\frac12 \mu(M) \bar{\Xi}(x)\Xi(x) 
+i\sqrt{2}\Lambda^2\phi(x) X(x) \right]. 
\eea
The SUSY transformation for $X$ and $\Xi$ is found as 
\bea
\delta_{\xi}X(x) & = & \bar{\xi}\left(\Xi(x) -i\sqrt{2}\frac{\Lambda^2}{\mu(M)^2}\psi(x)\right), \nono \\
\delta_{\xi}\Xi(x) & = & -\mu(M)\left(X(x) + i\sqrt{2}\frac{\Lambda^2}{\mu(M)^2}\phi(x)\right)\xi. 
\eea
After all, the total system (\ref{total_system}) can be reexpressed as 
\bea
\tilde{S}_{tot} & = & \tilde{S}_{FT} + S_{mess} + S_{MM}, \nono \\
\tilde{S}_{FT} & = & \int d^2 x \left[\frac12(\partial_\mu\phi)^2 +\frac12 \bar{\psi}\Slash{\partial} \psi 
  + \frac{\lambda^2}{2}\left(\phi^4 -\frac{m^2}{\lambda^2}\kappa\phi^2 + a^4\right) + \lambda\phi\bar{\psi}\psi \right]. 
  \label{rewrite_SFT}
\eea  

Now, we can regard $X$ and $\Xi$ as messenger fields sending the (SUSY breaking) effects from the ``hidden sector" 
($M$) to the ``visible sector" ($\phi$ and $\psi$). 
In ordinary phenomenological models, effects from hidden sectors induce soft SUSY breaking 
terms which {\em explicitly} break the SUSY in visible sectors. 
However, differently from that, in our model 
the ``hidden sector'' induces terms that cause the {\em spontaneous} SUSY breaking 
in the ``visible sector''. 
Let us set $\Lambda$ sufficiently heavier than $m$: $\Lambda \gg m$. Then, in $S_{mess}$ we can consider the 
situation that $X$ and $\Xi$ are originally dynamical whose hopping terms 
$\frac12(\partial_\mu X)^2$ and $\bar{\Xi}\Slash{\partial}\Xi$ can be neglected compared to the mass terms. 
After integrating out $M$, the mass squared $\vev{\mu(M)^2}$ behaves as 
\be
\vev{\mu(M)^2} = \left\{ \begin{array}{cc} \infty & \mbox{for finite $N$} \\
                                               \frac{2\Lambda^4}{m^2(1+\kappa)} & \mbox{for $N=\infty$}. 
                                               \end{array} \right. 
\ee
In the finite $N$ case, $X$ and $\Xi$ become infinitely heavy and 
cannot induce a $\phi^2$ term via the propagation of $X$. 
Thus, the potential in (\ref{rewrite_SFT}) does not change, 
and the SUSY is broken by instantons causing the tunneling 
between two minima of the double-well potential 
$\frac{\lambda^2}{2}(\phi^2-\frac{m^2}{2\lambda^2}\kappa)^2$. 
In the infinite volume limit, the effect vanishes and the SUSY is restored. 
On the other hand, in the $N=\infty$ case, their mass squared 
$\frac{2\Lambda^4}{m^2(1+\kappa)} (\gg m^2)$ is large but finite, 
which induces the $\phi^2$ term to change the potential to the single-well form 
$\frac{\lambda^2}{2}(\phi^2+\frac{m^2}{2\lambda^2})^2$.  
Thus, SUSY breaking takes place irrespective of finite or infinite volume. 

\section{O'Raifeartaigh Model Coupled to SUSY Matrix Model}
\label{sec:ORai}
\setcounter{equation}{0}
In this section, as the second example, we present the O'Raifeartaigh model coupled to
a Gaussian supersymmetric matrix model, where the matrix integration induces 
a term modifying the O'Raifeartaigh model to have a supersymmetric vacuum. 
When the size of the matrices $N$ becomes infinity, the position of the vacuum 
in the field space runs away to the infinity, and SUSY is broken.   

The O'Raifeartaigh model is defined in the four-dimensional space-time, and three chiral superfields 
$\Phi_0, \Phi_1, \Phi_2$ appear as its field contents. 
Using the Grassmann coordinates $\theta, \bar{\theta}$, they are written in terms of component fields 
as\footnote{We adopt the notation in Wess-Bagger's book.} 
\bea
\Phi_a(y) & = & \phi_a(y) + \sqrt{2} \theta \psi_a(y) + \theta\theta F_a(y) \qquad (a=0, 1, 2), \nono \\
y^\mu & = & x^\mu + i\theta\sigma^\mu \bar{\theta}. 
\eea
The action is given by 
\bea
S_{OR} & = & \int d^4x \, {\cal L}_{OR}, \nonumber \\
{\cal L}_{OR} & = & \int d^2\theta d^2\bar{\theta}\, K + \int d^2\theta \, W + \int d^2\bar{\theta} \, \bar{W}, 
\label{SOR}
\eea
where the K\"ahler potential $K$ has the canonical form: 
\be
K=\Phi_0\bar{\Phi}_0 + \Phi_1\bar{\Phi}_1 + \Phi_2\bar{\Phi}_2, 
\ee
and the superpotential is taken as 
\be
W=\lambda \Phi_0 + \mu \Phi_1\Phi_2 + g \Phi_0\Phi_1^2. 
\label{ORaiW}
\ee
$\lambda$, $\mu$, $g$ are coupling constants. 
After integrating out the auxiliary fields $F_a$, we find the scalar potential  
\be
V = |\lambda + g\phi_1^2|^2 + |\mu\phi_2 + 2g\phi_0\phi_1|^2 + |\mu\phi_1|^2.
\label{ORscpot}
\ee
Because $V>0$ for all $\phi_0, \phi_1, \phi_2$ as long as $\lambda\neq 0$, 
the SUSY is spontaneously broken. 

Here, we introduce the matrix-valued (anti)chiral supervariables, 
which are constant as a function of $y^\mu$ ($y^{\mu\dagger}$): 
\be
{\cal M} = M + \sqrt{2}\theta\chi + \theta\theta F_M, \qquad 
\bar{\cal M} = \bar{M} + \sqrt{2}\bar{\theta}\bar{\chi} + \bar{\theta}\bar{\theta} \bar{F}_M
\ee
with $M, \bar{M}$, $\chi_{\alpha}, \bar{\chi}^{\dot{\alpha}}$, $F_M, \bar{F}_M$ 
being $N\times N$ complex matrices\footnote
{This kind of variables was used in the large-$N$ twisted reduced model 
of the four-dimensional ${\cal N}=1$ supersymmetric field theory~\cite{Kawai:2003yf}.}. 
As the simplest case, let us choose the Gaussian action for the MM sector: 
\be
S_{MM} = \tr \left[\int d^2\theta d^2\bar{\theta} \, {\cal M} \bar{\cal M}
+ \frac{m}{2} \int d^2\theta \, {\cal M}^2 + \frac{\bar{m}}{2} \int d^2\bar{\theta} \, \bar{\cal M}^2\right]. 
\label{S_MM_ch}
\ee
Note that the same superspace coordinates $\theta, \bar{\theta}$ 
are used in the O'Raifeartaigh model and in the matrix model. 
Also, we take the interactions between the O'Raifeartaigh model and the matrix model as 
\be
S_{int} = f\int d^2\theta \left(\frac{1}{N}\tr {\cal M} \int d^4x \, \Phi_0 \right)+
\bar{f}\int d^2\bar{\theta} \left(\frac{1}{N}\tr \bar{\cal M} \int d^4x \, \bar{\Phi}_0\right). 
\label{int_ORai}
\ee
Interestingly, the interaction terms can be obtained by promoting the coupling of the O'Raifeartaigh model $\lambda$ 
to a dynamical variable in the MM sector as 
\be
\lambda \to \tilde{\lambda}({\cal M}) \equiv \lambda + f \frac{1}{N}\tr {\cal M}.
\label{tilde_lambda}
\ee

After the Wick rotation, the total Euclidean action becomes 
\be
S_{tot} = -S_{OR} - S_{MM} - S_{int}, 
\ee 
where $x^0$ appearing in ${\cal L}_{OR}$ is understood as $-ix^0$ by the Wick rotation\footnote{
Although the Gaussian matrix model here can consistently couple to the field theory sector also in the Minkowski 
space, matrix models with self-interactions are not allowed to couple in the 
Minkowski space. In the case of the Euclid space, they have a consistent coupling. 
For the explanation, see the last paragraph in appendix B.}.   
For the integrals of $F_a, \bar{F}_a$ fields, the integration contours are taken parallel to the imaginary axis 
so that the integrals are well-defined. 
The total action has the four-dimensional ${\cal N}=1$ SUSY corresponding to a shift of the supercoordinates 
$\theta, \bar{\theta}$. 

We compute the induced action $\Gamma$ obtained by integrating out matrix variables: 
\be
e^{-\Gamma} = \int  (d{\cal M}\, d\bar{\cal M})\, e^{-S_{tot}}. 
\ee
To see SUSY breaking, it is relevant to investigate the potential part of $\Gamma$, 
which depends only on the constant parts of the chiral superfields $\Phi_a$. 
Expressing the constant parts as 
\be
\Phi_a^{(0)} = \frac{1}{V_{4D}}\int d^4x \, \Phi_a \qquad (a=0,1,2)
\label{const_mode}
\ee
with $V_{4D}$ the volume of the space-time, 
the potential part $\Gamma^{(0)}$ is given by 
\bea
\frac{\Gamma^{(0)}}{V_{4D}}& = & -\int d^2\theta d^2\bar{\theta} \, {\cal K} -\int d^2\theta \, {\cal W} 
-\int d^2\bar{\theta} \, \bar{\cal W}, \nonumber \\
{\cal K} & = & \left(1+\frac{f\bar{f}V_{4D}}{Nm\bar{m}}\right)\Phi_0^{(0)}\bar{\Phi}_0^{(0)}
+ \Phi_1^{(0)}\bar{\Phi}_1^{(0)}+\Phi_2^{(0)}\bar{\Phi}_2^{(0)}, \nonumber \\
{\cal W} & = & -\frac{f^2V_{4D}}{2Nm}\left(\Phi_0^{(0)}\right)^2 +\lambda \Phi_0^{(0)} 
+\mu \Phi_1^{(0)}\Phi_2^{(0)} + g\Phi_0^{(0)}\left(\Phi_1^{(0)}\right)^2. 
\label{inducedKW}
\eea
The effect of the matrix integrals changes the weight of the $\Phi^{(0)}_0\bar{\Phi}^{(0)}_0$ in the K\"ahler potential      
and more importantly generates the $\left(\Phi_0^{(0)}\right)^2$ term in the superpotential. 
After rescaling $\Phi_0^{(0)}$ as 
\be
X = \sqrt{1+\frac{f\bar{f}V_{4D}}{Nm\bar{m}}} \,  \Phi_0^{(0)} 
\ee
to make the K\"ahler potential canonical, the superpotential 
\be
{\cal W} = -\frac12 \epsilon X^2 + \lambda'X +\mu \Phi_1^{(0)}\Phi_2^{(0)} + g'X\left(\Phi_1^{(0)}\right)^2
\label{inducedW}
\ee 
with 
\bea
 & & \epsilon \equiv  \frac{f^2V_{4D}}{Nm} \left(1+\frac{f\bar{f}V_{4D}}{Nm\bar{m}}\right)^{-1}, \nono \\
 & & \lambda' \equiv  \lambda \left(1+\frac{f\bar{f}V_{4D}}{Nm\bar{m}}\right)^{-1/2}, \qquad 
 g' \equiv  g 
\left(1+\frac{f\bar{f}V_{4D}}{Nm\bar{m}}\right)^{-1/2}
\eea
turns out to have the supersymmetric vacuum at 
\be
\vev{X} = \frac{\lambda'}{\epsilon}, \qquad \vev{\Phi_1}=\vev{\Phi_2} =0. 
\label{SUSYvac}
\ee

Considering the case where $m, \bar{m}, f, \bar{f}$ are finite and $N\gg V_{4D}$, $\epsilon$ is small and 
$\lambda' \approx  \lambda$. 
The supersymmetric vacuum (\ref{SUSYvac}) exists, no matter how small $\epsilon$ is as long as it is nonzero.  
However, for $N=\infty$, $\epsilon$ vanishes, and the vacuum runs away to the infinity in the field space and disappears. 
It means that the supersymmetry is broken. 
Indeed, the superpotential (\ref{inducedW}) is reduced 
to (\ref{ORaiW}) in the original 
O'Raifeartaigh model, where the SUSY is broken as we saw. 
Thus, the model realizes our desirable situation that the SUSY is preserved for $N$ finite, 
but becomes broken at $N=\infty$. 
Correspondingly, the Witten index is one for any finite $N$, but the value jumps to zero at $N=\infty$ due to the change 
modifying the asymptotic behavior of the potential. 

There are several comments in order. 
Firstly, we took $m, \bar{m}, f, \bar{f}$ to be ${\cal O}(1)$ quantities in the above. 
It is quite plausible from the point of view of the naturalness, 
where we should start with all the coupling constants being ${\cal O}(1)$. 
In fact, $m, \bar{m}={\cal O}(1)$ means that the fluctuation of each component of ${\cal M}, \bar{\cal M}$ is ${\cal O}(1)$, and 
$f, \bar{f}={\cal O}(1)$ can be interpreted as 
$\tilde{\lambda}({\cal M})$ in (\ref{tilde_lambda}) and its complex conjugate being ${\cal O}(1)$, 
which are the ``coupling constants'' for $\Phi_0, \bar{\Phi}_0$. 

Secondly, we considered the interaction terms (\ref{int_ORai}) to be local with respect to $\theta, \bar{\theta}$ 
with ${\cal M}$ and $\Phi_0$ expanded by the same  $\theta, \bar{\theta}$. 
This is crucial to induce $\left(\Phi_0^{(0)}\right)^2$ term in the superpotential ${\cal W}$. 
If we worked out with the ``nonlocal interaction terms'': 
\be
f\left(\int d^2\theta \frac{1}{N}\tr {\cal M} \right) \left(\int d^2\theta \int d^4x \, \Phi_0 \right)+
\bar{f}\left(\int d^2\bar{\theta} \frac{1}{N}\tr \bar{\cal M} \right) \left(\int d^2\bar{\theta} \int d^4x \, \bar{\Phi}_0\right)
\ee
instead of (\ref{int_ORai}), the induced terms would take the form of 
$$\left(\int d^2\theta \, \Phi_0^{(0)}\right)^2=\left(F_0^{(0)}\right)^2, 
\qquad \left(\int d^2\bar{\theta} \, \bar{\Phi}_0^{(0)}\right)^2=\left(\bar{F}_0^{(0)}\right)^2, 
$$
which could not yield a supersymmetric vacuum.  

Thirdly, this kind of modification of the superpotential (\ref{inducedW}) is discussed in Ref.~\cite{Dine:2006gm}. 
Here, the smallness of the (dimensionless) parameter $\epsilon/\mu$ is naturally achieved 
by the large matrix degrees of freedom $N$. 
It presents a new viewpoint on the dynamical origin of the small $\epsilon/\mu$ different from the discussion 
in \cite{Dine:2006gm}, where the strong gauge dynamics of the hidden sector is considered to explain the smallness. 
Also, in \cite{Intriligator:2007cp}, the modification of the superpotential 
by adding $-\frac12\epsilon \Phi_2^2$ instead of 
$-\frac12\epsilon X^2$ is discussed. 
By replacing $\Phi_0, \bar{\Phi}_0$ with $\Phi_2, \bar{\Phi}_2$ in (\ref{int_ORai}) and repeating the same computation, 
it is possible to reproduce that kind of modification in our setup. 

Finally, from the viewpoint of the dynamical variable $\tilde{\lambda}({\cal M})$ in (\ref{tilde_lambda}), 
we can say that introducing the fluctuation of the coupling $\lambda$ along the $\theta$-direction in the O'Raifeartaigh model 
leads the $\left(\Phi_0^{(0)}\right)^2$ term in the superpotential, and makes restored the SUSY which was broken in the original 
O'Raifeartaigh model.      
It should be also noted that the integration over matrix variables 
is essential in yielding the $\left(\Phi_0^{(0)}\right)^2$ term.

\section{Interacting Matrices and Metastable Vacua}
\label{sec:meta}
\setcounter{equation}{0}
Here, we discuss two extensions of the setting in the previous section, where the matrix model sector was Gaussian and 
the sector of the O'Raifeartaigh model was treated at the classical level. 
One is to introduce cubic interaction terms in the matrix model and analyze the SUSY breaking of the induced potential 
after integrating the matrices in the perturbative expansion. 
The other is to compute the one-loop effective potential obtained by integrating out heavy degrees of freedom 
($\Phi_1$ and $\Phi_2$) of
the induced model in the previous section, and to see that a 
metastable vacuum appears 
near the origin of the field space of $\Phi_0$.  

\subsection{Cubic Interactions of Matrices} 
We add the cubic interaction terms 
\be
\Delta S_{MM} =\tr \left[\frac{h}{3} \int d^2\theta \, {\cal M}^3 
+ \frac{\bar{h}}{3} \int d^2\bar{\theta}\, \bar{\cal M}^3 \right]
\ee
to the matrix model action $S_{MM}$ (\ref{S_MM_ch}), 
and repeat similar computation as in the previous section 
with the perturbative treatment of $\Delta S_{MM}$ considering the case 
$|h|, |\bar{h}| \ll 1/\sqrt{N}$. 
As a result of the matrix integration up to the second order with respect to $h, \bar{h}$, 
we obtain the following induced K\"ahler potential and superpotential: 
\bea
{\cal K} & = & \left\{1-\left(1-2\frac{h\bar{h}N}{m^2\bar{m}^2}\right)\frac{f\bar{f}V_{4D}}{Nm\bar{m}}\right\}
\tilde{\Phi}_0^{(0)}\bar{\tilde{\Phi}}_0^{(0)} 
+\frac{h\bar{h}(f\bar{f})^2V_{4D}^3}{N^3m^3\bar{m}^3}\left(\tilde{\Phi}_0^{(0)}\bar{\tilde{\Phi}}_0^{(0)}\right)^2 \nonumber \\
 & & + \Phi_1^{(0)}\bar{\Phi}_1^{(0)} + \Phi_2^{(0)}\bar{\Phi}_2^{(0)} 
 +\frac{g\bar{f}}{\bar{m}f}\bar{\tilde{\Phi}}_0^{(0)}\left(\Phi_1^{(0)}\right)^2 
 +\frac{\bar{g}f}{m\bar{f}}\tilde{\Phi}_0^{(0)}\left(\bar{\Phi}_1^{(0)}\right)^2 \nono \\
 & & + {\cal O}(h^2\bar{h}, h\bar{h}^2), 
\label{inducedK2}\\
 {\cal W} & = & -\frac{f^2V_{4D}}{2Nm}\left(\tilde{\Phi}_0^{(0)}\right)^2 + \lambda \tilde{\Phi}_0^{(0)} +\mu \Phi_1^{(0)}\Phi_2^{(0)} 
 +g\tilde{\Phi}_0^{(0)}\left(\Phi_1^{(0)}\right)^2 \nonumber \\
 & & -\frac{hf^3V_{4D}^2}{3N^2m^3}\left(\tilde{\Phi}_0^{(0)}\right)^3 - \frac{h^2f^4V_{4D}^3}{2N^3m^5}\left(\tilde{\Phi}_0^{(0)}\right)^4 
+ {\cal O}(h^3), 
\label{inducedW2}
\eea 
where we used the shifted variables 
\be
\tilde{\Phi}_0^{(0)} \equiv \Phi_0^{(0)} -\frac{\bar{f}}{\bar{m}f}\bar{F}_0^{(0)}, \qquad 
\bar{\tilde{\Phi}}_0^{(0)} \equiv \bar{\Phi}_0^{(0)} -\frac{f}{m\bar{f}}F_0^{(0)}. 
\ee
(See appendix B for the computational details.) 

Rescaling the shifted fields to have a canonical normalization of $\tilde{\Phi}_0^{(0)}\bar{\tilde{\Phi}}_0^{(0)}$ in 
${\cal K}$: 
\be 
X = \sqrt{1-\left(1-2\frac{h\bar{h}N}{m^2\bar{m}^2}\right)\frac{f\bar{f}V_{4D}}{Nm\bar{m}}} \, \tilde{\Phi}_0^{(0)} , 
\ee
we find SUSY minima at 
\be
\vev{X} = {\cal O}\left(\frac{N}{V_{4D}}\right), \qquad \vev{\Phi_1^{(0)}} = \vev{\Phi_2^{(0)}} = 0. 
\label{SUSYvac2}
\ee
Generically, the number of the SUSY minima is equal to the degrees of $X$ in $\frac{\partial {\cal W}}{\partial X}$, 
and all of them are located at points with the distance ${\cal O}\left(\left|\frac{N}{V_{4D}}\right|\right)$ from the origin. 
For the case up to the order ${\cal O}(h^2)$ considered here, we find the three minima. 
In the limit $h, \bar{h}\to 0$, two of them run to the infinity and disappear, and the remaining one gives the minimum (\ref{SUSYvac}) 
reproducing the situation in the previous section. 
For the large-$N$ limit satisfying $N\gg V_{4D}$, all the minima (\ref{SUSYvac2}) move to the infinity, 
and the SUSY is 
broken. 

Hence, our mechanism for the SUSY breaking at $N=\infty$ in the previous section works not only 
for the Gaussian supersymmetric matrix model action 
but also for the action including the cubic interactions. 
Furthermore, it is expected to do so for more general supersymmetric matrix model actions at least perturbatively.

\subsection{One-loop Effects and Metastable Vacua}
The scalar potential corresponding to the induced K\"ahler potential (\ref{inducedW}) reads 
\be
V = |\lambda'-\epsilon \phi_X + g' \phi_1^2|^2 + |\mu \phi_2+2g'\phi_X\phi_1|^2 + |\mu\phi_1|^2,  
\ee
where $\phi_X$ denotes the lowest component of $X$. 
Expanding this with respect to $\phi_1, \phi_2$ around the origin, we obtain 
\be
V  =   |\lambda'-\epsilon \phi_X|^2 + \frac12 B^\dagger M_0^2 B + \cdots, 
\label{scalarpot}
\ee 
with
\be
B =  \left(  \begin{array}{c} \phi_1 \\ \phi_2 \\ \bar{\phi}_1 \\ \bar{\phi}_2 \end{array} \right), \qquad 
B^\dagger = \left( \bar{\phi}_1, \bar{\phi}_2, \phi_1, \phi_2 \right), 
\ee
\be
M_0^2 \equiv \left[\begin{array}{cccc} |\mu|^2 + |2g'\phi_X|^2 & 2\bar{g}'\mu \bar{\phi}_X & 2(\lambda'-\epsilon \phi_X)\bar{g}' & 0 \\
    2g'\bar{\mu} \phi_X & |\mu|^2 & 0 & 0 \\
    2(\bar{\lambda}' - \bar{\epsilon}\bar{\phi}_X)g' & 0 & |\mu|^2 + |2g'\phi_X|^2 & 2g'\bar{\mu} \phi_X \\
    0 & 0 & 2\bar{g}'\mu \bar{\phi}_X & |\mu|^2 \end{array} \right], 
\ee    
and the ellipsis meaning contributions from higher than the quadratic orders of $\phi_1, \phi_2$. 
The eigenvalues of the scalar (mass)$^2$ matrix $M_0^2$ are given by 
\bea
\left. \begin{array}{c} m_{B1}^2 \\ m_{B2}^2 \end{array} \right\} & = & 
|\mu|^2 + 2|g'\phi_X|^2 + |g'(\lambda'-\epsilon \phi_X)| \pm \sqrt{\left\{2|g'\phi_X|^2 + |g'(\lambda-\epsilon \phi_X)|\right\}^2 
+ 4|g'\mu \phi_X|^2} , \nono \\
\left. \begin{array}{c} m_{B3}^2 \\ m_{B4}^2 \end{array} \right\} & = & 
|\mu|^2 + 2|g'\phi_X|^2 - |g'(\lambda'-\epsilon \phi_X)| \pm \sqrt{\left\{2|g'\phi_X|^2 - |g'(\lambda-\epsilon \phi_X)|\right\}^2 
+ 4|g'\mu \phi_X|^2}, \nono \\
 & &  
\label{mB} 
\eea
from which it can be seen that in order for $M_0^2$ 
to be positive definite, the condition 
\be
\left|\frac{2g'}{\mu^2}(\lambda'-\epsilon \phi_X)\right| < 1
\label{nontachyonic}
\ee
must be met. 
When it is so, the potential at 
\be
\phi_1=\phi_2 =0, \qquad \phi_X = {\rm arbitrary}
\label{Xarb}
\ee
is stable 
with respect to the fluctuations of $\phi_1$ and $\phi_2$, but not 
stable with respect to $\phi_X$ except the point (\ref{SUSYvac}), 
as easily seen from the first term of (\ref{scalarpot}). 
However, for infinitesimally small $\epsilon$, it does not change much  as $\phi_X$ moves. 
In the limit $\epsilon \to 0$ where the original O'Raifeartaigh model 
is recovered, 
the classical potential at the point (\ref{Xarb}) does not depend on $\phi_X$, and 
$\phi_X$ can be regarded as a classical moduli. 
In refs.~\cite{Intriligator:2007cp}, it is called ``pseudomoduli'' since the one-loop calculation for the fluctuations with respect to 
$\Phi_1, \Phi_2$ induces $X$-dependence lifting the degeneracy 
of classical vacua. 
In our case, although the degeneracy for $\phi_X$ is already lifted 
due to $\epsilon$ 
and the potential is not stable near $\phi_X\sim 0$, 
we will see that the one-loop contribution induces new $\phi_X$-dependence, 
which stabilizes the potential near $\phi_X\sim 0$ yielding a metastable vacuum there. 
In this subsection, we consider the case of small $g'$ 
to trust the one-loop approximation.  

{}From the superpotential (\ref{inducedW}), the fermion mass terms at (\ref{Xarb}) read\footnote{
$\psi_X, \bar{\psi}_X$ denote the superpartners of $\phi_X, \bar{\phi}_X$, respectively.} 
\be
\frac12\epsilon \psi_X\psi_X + \frac12(\psi_1, \psi_2)M_{1/2}\left(\begin{array}{c} \psi_1 \\ \psi_2 \end{array} \right) 
+ \frac12\bar{\epsilon} \bar{\psi}_X\bar{\psi}_X + \frac12(\bar{\psi}_1, \bar{\psi}_2)M_{1/2}^\dagger
\left(\begin{array}{c} \bar{\psi}_1 \\ \bar{\psi}_2 \end{array} \right) 
\label{Fmassterm}
\ee
with the mass matrices 
\be
M_{1/2} = \left[\begin{array}{cc} 2g'\phi_X & \mu \\ \mu & 0 \end{array} \right], \qquad 
M_{1/2}^\dagger = \left[\begin{array}{cc} 2\bar{g}'\bar{\phi}_X & \bar{\mu} \\ \bar{\mu} & 0 \end{array} \right]. 
\ee
The (mass)$^2$ eigenvalues of the matrix $M_{1/2}M_{1/2}^\dagger$ are given by 
\be
\left. \begin{array}{c} m_{F1}^2 \\ m_{F2}^2 \end{array} \right\} = \left(|g'\phi_X| \pm \sqrt{|\mu|^2 + |g'\phi_X|^2}\right)^2. 
\label{mF}
\ee

The one-loop integrals with respect to $\phi_{1, 2}$ and $\psi_{1, 2}$ induce the potential per unit volume 
(the Coleman-Weinberg potential): 
\be
V_{\rm CW}  =   \int^\Lambda \frac{d^4p}{(2\pi)^4}
\left\{ \sum_{i=1}^4 \frac12 \log (p^2 + m_{Bi}^2) -\sum_{i=1}^2 \log (p^2 +m_{Fi}^2)\right\}, 
\ee
where 
$\Lambda$ represents the ultraviolet cutoff for the radial part of the momentum 
$p\equiv \sqrt{p_\mu p_\mu}$, and note that each of the fermion (mass)$^2$ has the degeneracy two. 
Due to the sum rule 
\be
\sum_{i=1}^4 m_{Bi}^2 = 2\sum_{i=1}^2 m_{Fi}^2, 
\ee
divergent terms of the order ${\cal O}(\Lambda^2)$ or higher are cancelled, and thus we find 
\be
V_{\rm CW} = \frac{1}{64\pi^2}\left[ \sum_{i=1}^4 m_{Bi}^4 \log \frac{m_{Bi}^2}{M_{\rm cutoff}^2} 
-2\sum_{i=1}^2 m_{Fi}^4\log \frac{m_{Fi}^2}{M_{\rm cutoff}^2}\right].  
\ee
Here, we put $M_{\rm cutoff} \equiv e^{1/4} \Lambda$ for notational simplicity. 
Explicitly, 
\bea
V_{\rm CW} & = & \frac{|\mu|^4}{64\pi^2}\left[8z^2\log\frac{|\mu|^2}{M_{\rm cutoff}^2} + h(x,z)+h(x,-z)-2h(x,0) \right],\nono \\
h(x,z) &\equiv & \left(1+x+z-\sqrt{(x+z)^2+2x}\right)^2\log(1+2z) \nono \\
 & & +4(1+x+z)\sqrt{(x+z)^2+2x} \, \log\left(1+x+z+\sqrt{(x+z)^2+2x}\right), 
\label{explicit_CW}
\eea
where
\be
x=2\left|\frac{g'\phi_X}{\mu}\right|^2, \qquad z=\left|(\lambda'-\epsilon \phi_X)\frac{g'}{\mu^2}\right|. 
\label{variables_xz}
\ee
Together with the classical potential $V_{\rm tree} = |\lambda'-\epsilon \phi_X|^2$, 
we obtain the effective potential up to the one-loop order 
\be
V_{\rm eff}^{(1)} = V_{\rm tree} + V_{\rm CW},  
\label{eff_pot}
\ee
whose explicit form for small $|\phi_X|$ is 
\be
V_{\rm eff}^{(1)} = V_0 - \frac{R}{2} \left(\frac{\epsilon}{\lambda'}\phi_X+\frac{\bar{\epsilon}}{\bar{\lambda}'}\bar{\phi}_X\right) 
+ m_X^2 |\phi_X|^2 + {\cal O}(|\phi_X|^3), 
\label{Veff}
\ee 
where the constants are 
\bea
V_0 & = & |\lambda'|^2\left[1+ \frac{|g'|^2}{8\pi^2}\left(\log \frac{|\mu|^2}{M_{\rm cutoff}^2} +\frac32 + v(y)\right) 
+{\cal O}(|g'|^4) \right], \nono \\
v(y) & \equiv & \frac{(1+y)^2}{2y^2}\log (1+y) + \frac{(1-y)^2}{2y^2}\log (1-y) -\frac32, \nono \\
y & \equiv & \left|\frac{2\lambda' g'}{\mu^2}\right|, \nono \\
R & \equiv & 2V_0 + \frac{|\lambda'g'|^2}{8\pi^2}\left(1-\frac{1+y}{y^2}\log (1+y) -\frac{1-y}{y^2}\log (1-y)\right) \nono \\
 & = & 2|\lambda'|^2 +\frac{|\lambda'g'|^2}{8\pi^2}\left[2\log\frac{|\mu|^2}{M_{\rm cutoff}^2}+1
+\frac{1+y}{y}\log (1+y) -\frac{1-y}{y}\log (1-y)\right], \nono \\
m_X^2 & \equiv & \frac{1}{2\pi^2}\left|\frac{\lambda'^2g'^4}{\mu^2}\right|\nu (y) + {\cal O}(|g'|^4), \nono \\
\nu (y) & \equiv & \frac{1}{y^3}\left[(1+y)^2\log (1+y) -(1-y)^2\log(1-y) -2y\right]. 
\eea
Assuming $|\epsilon|$ infinitesimally small, we took $\epsilon, \bar{\epsilon}$-dependent terms up to the linear order. 
We find that $V_{\rm eff}^{(1)}$ has a local minimum near the origin at 
\be
\phi_{X\, \rm{min}} = \frac{R}{2m_X^2}\frac{\bar{\epsilon}}{\bar{\lambda}'} 
\ee
with the height $V_0$. 

The local minimum can be regarded as a metastable vacuum which eventually decays to the true vacuum at 
$\phi_X = \frac{\lambda'}{\epsilon}$. 
We estimate the lifetime of the metastable vacuum 
by computing the bounce action for an approximated potential of the shape of a triangle as in~\cite{Duncan:1992ai}. 
For details, see appendix~\ref{app:lifetime}. 
It is estimated to be of the order  
\be
{\cal O}\left(\exp \left\{|\epsilon|^{-4}\right\}\right)= {\cal O}\left(\exp\left\{\left(\frac{N}{V_{4D}}\right)^4\right\}\right),
\ee
which is sufficiently long for infinitesimal $|\epsilon|$ or large $N$. 

The effective potential $V_{\rm eff}^{(1)}$ depends only on the constant mode $\phi_X$, and 
it can be regarded as a nonlocal potential of $\phi_0(x)$. 
In general, it is not clear if the method of \cite{Duncan:1992ai} can be applied to the nonlocal potential. 
According to the last paragraph of appendix~\ref{app:lifetime}, 
however, at least in our case, 
it is valid if the condition 
\be
N\gg V_{4D}^{5/4}
\ee
is satisfied.  
It is slightly stronger than $N\gg V_{4D}$ which appeared in the previous section, but 
it gives no obstruction to take the large-$N$ limit.

\section{Outlook}
\label{sec:outlook}
\setcounter{equation}{0}
Motivated to discuss SUSY breaking in matrix models proposed for the nonperturbative formulation of 
superstring theory, in this paper we have presented some concrete models of the form of supersymmetric field theory 
coupled to some matrix model, where the SUSY of the total system is preserved for any finite 
$N$ (the size of the matrices appearing in the matrix model), but spontaneously broken at $N=\infty$. 

The first model is Witten's simple two-dimensional supersymmetric model coupled to a hermitian one-matrix model, 
in which the matrix integrals induce a lift of the field theory potential at $N=\infty$. 
This mechanism is applicable to other models not only to the two-dimensional model. 
For example, we can use it to the O'Raifeartaigh model by considering the following coupling: 
\be
S_{tot} =  -\widehat{S}_{OR} + S_{MM}, 
\ee
where $\widehat{S}_{OR}$ is given by (\ref{SOR}) whose superpotential (\ref{ORaiW}) is replaced with 
\bea
\widehat{W} & = & \widehat{\lambda}(M) \Phi_0 + \mu \Phi_1\Phi_2 + g \Phi_0\Phi_1^2, \nono \\
\widehat{\lambda}(M) & \equiv & \lambda \, \frac{\epsilon^2}{(\frac{1}{N}\tr M)^2 + \epsilon^2}, 
\label{hatW}
\eea
and $S_{MM}$ is the one-matrix model potential (\ref{MMpot}). By a similar argument as in section 2, 
as a result of the matrix integration, we see
\be
\vev{\widehat{\lambda}(M)} = \left\{ \begin{array}{cl} 0 & \mbox{for $N$: finite} \\
                                \lambda & \mbox{for $N=\infty$} \end{array} \right.                
\ee  
with the higher-order contributions of $\vev{\widehat{\lambda}(M)^{k}}$ ($k\ge 2$) being irrelevant. 
For finite $N$ case, the scalar potential becomes (\ref{ORscpot}) with $\lambda$ set to zero, yielding 
the supersymmetric minimum $\phi_1=\phi_2 =0$, $\phi_0 = \mbox{arbitrary}$. 
At $N=\infty$, however, the system is reduced 
to the ordinary O'Raifeartaigh model causing the SUSY breaking. 

The mechanism can be applied also to a four-dimensional supersymmetric gauge theory, 
whose gauge group contains the U(1) factor. 
We incorporate the coupling between the gauge theory and the one-matrix model through the Fayet-Iliopoulos (FI) term: 
\be
2\widehat{\kappa}(M) \int d^4x \, d^2\theta d^2\bar{\theta} \, \tr V, 
\ee
where $V$ is the vector superfield, and the FI parameter $\widehat{\kappa}(M)$ is given by 
\be
\widehat{\kappa}(M)  \equiv  \kappa \, \frac{\epsilon^2}{(\frac{1}{N}\tr M)^2 + \epsilon^2}.  
\ee  
After the matrix integrals, the effect of the FI term emerges 
only at $N=\infty$, breaking the supersymmetry and the U(1) 
gauge symmetry. 

The second model we have discussed is the O'Raifeartaigh model coupled to the supersymmetric matrix model. 
Similarly to (\ref{hatW}), the coupling between the two sectors is introduced 
by changing the coupling constant $\lambda$ in the O'Raifeartaigh superpotential to $\tilde{\lambda}({\cal M})$ 
as in (\ref{tilde_lambda}). 
However, differently from the first model and its variants discussed above, 
the matrix variables ${\cal M}, \bar{\cal M}$ have $\theta, \bar{\theta}$-dependence. 
The matrix integration induces the $-\frac12\epsilon \left(\Phi_0^{(0)}\right)^2$ term in the superpotential yielding the supersymmetric vacuum 
at $\vev{\phi_0} = {\cal O}\left(\frac{1}{\epsilon}\right)$. Since the value of $\epsilon$ decreases to zero as $N$ becomes larger, 
the supersymmetric minimum runs to the infinity and disappears in the large-$N$ limit. 

This mechanism can be applied to the supersymmetry breaking in supersymmetric QCD (SQCD), 
because the O'Raifeartaigh type interaction appears in the low energy effective action of SQCD.  
The starting microscopic theory is SU($N_c$) gauge theory 
with $N$ flavors. As discussed in \cite{Seiberg:1994pq}, in the case $N >N_c$, 
the dual description of the gauge group SU($n$) ($n=N-N_c$) is useful for the strongly coupled low energy region. 
The superpotential can be expressed as 
\be
W = h \,\tr (\Phi\varphi\tilde{\varphi}^T) + \tr (f\Phi), 
\label{WSQCD}
\ee
where $\Phi$, $\varphi$, $\tilde{\varphi}$ are matrix-valued chiral superfields with the sizes $N \times N$, $N\times n$, 
$N\times n$, respectively, and 
$f$ is a coupling matrix of the size $N\times N$. 
The K\"ahler potential has the canonical form. 
In terms of the original $N$ quarks $Q$ and $N$ anti-quarks $\tilde{Q}$, the fields describing mesonic 
and baryonic degrees of freedom are 
\be
\tilde{Q}Q^T \sim \Phi, \qquad Q^{N_c} \sim \varphi^n, \qquad \tilde{Q}^{N_c} \sim \tilde{\varphi}^n. 
\ee
When $\mbox{rank}\,f >n$, the condition for the supersymmetric vacua 
\be
0 = h\varphi\tilde{\varphi}^T + f
\ee
can not be satisfied, and the SUSY is broken.  

Let us consider the model coupled to the supersymmetric matrix model (\ref{S_MM_ch}) through changing $f$ to 
\be
f \to \tilde{f}_1({\cal M}) \equiv f + \frac{\gamma}{N}{\cal M}. 
\label{tilde_f}
\ee
The matrix integrals induce the term of the form $-\frac12 \epsilon \left(\Phi^{(0)}\right)^2$ in the superpotential\footnote{
$\Phi^{(0)}$ is the constant mode of $\Phi$ as in the notation (\ref{const_mode}).}, and similarly as above,  
the SUSY is restored for any finite $N$, but gets broken for $N=\infty$ 
at the classical level\footnote{Here, we do not take into 
account the effects of the Vandermonde determinant arising from the path-integral measure of the field theory sector.}. 
Note that if we consider another coupling 
\be
f \to \tilde{f}_2({\cal M}) \equiv f + \frac{\gamma}{N}\, \tr {\cal M} 
\label{tilde_f2}
\ee
instead of (\ref{tilde_f}), after the matrix integration, the condition for the supersymmetric vacua becomes 
\be
 0= \left(h\varphi\tilde{\varphi}^T + f \right)_{ij} - \delta_{ij}\frac{\gamma^2V_{4D}}{N}\, \tr \Phi^{(0)},    
\ee
which can not be satisfied 
as long as $n<\mbox{rank}\,f< N$. 
Then, the SUSY is broken not only for $N=\infty$ 
but also for some finite $N$. 
Thus, the coupling (\ref{tilde_f}) is necessary to realize our desirable SUSY breaking rather than (\ref{tilde_f2}). 

As we saw, it is clear that the mechanisms presented here are applicable to many models for SUSY breaking. 
Also, the large degrees of freedom associated with $N$ 
naturally explain the smallness of some parameters, for example $\epsilon$ in 
the O'Raifeartaigh model case. 
In fact, the possibility that a large amount of the degrees of freedom explains some naturalness issues 
is discussed in the contexts of the inflation~\cite{Dimopoulos:2005ac} 
and the hierarchy between the Planck scale and the baryon mass scale~\cite{Dvali:2007hz}. 
The application of our mechanism to such problems would be intriguing from the phenomenological viewpoint, 
and furthermore would give a new insight 
into considering the realization of natural phenomenological models 
starting with matrix model formulations of superstrings.  

In the models we have discussed so far, the dynamical aspects of the MM sector seem not very important to 
the SUSY breaking in the FT sector. 
It is interesting to construct the examples, in which the dynamical symmetry breaking of the MM sector, 
taking place only at large $N$, directly causes the SUSY breaking in the FT sector.      
We had considered the two-matrix model
\be
S_{MM} = N \tr \left[\frac12 A^2 -\frac{g}{3}A^3 + \frac12 B^2 -\frac{g}{3}B^3 -cAB\right]
\ee
and the spin operator on random surfaces $\tr(A-B)$ as the coupling to the FT sector $a^2$ in (\ref{pro_FT}), 
intending the spontaneous magnetization in the matrix model 
in the large-$N$ limit to induce the SUSY breaking. 
But it does not work, because the magnetization can not be seen 
in the planar one- or two-point functions on the random surfaces. 
Nevertheless, this kind of challenge would reveal new interesting aspects of matrix models.

\section*{Acknowledgements}
We would like to thank Satoshi Iso, Hiroshi Suzuki and Kazunori Takenaga
for useful discussions. 
This work is benefited by the SAKURA project exchanging researchers between France and Japan; T.K. would like to thank CEA/Saclay for hospitality while this work was in progress. 
The authors thank the Yukawa Institute for Theoretical Physics at Kyoto University. Discussions during the YITP workshop YITP-W-07-05 on ``String Theory and Quantum Field Theory'' were useful to complete this work.

\appendix
\section{$\vev{a_0^{2k}}\to 0$ for Finite $N$}
\label{app:proto}
\setcounter{equation}{0}
Let us consider a hermitian one-matrix model defined by the action 
\be
S_{MM} = N \tr\, V(M), \qquad V(-M) = V(M). 
\ee
The potential $V(x)$ is bounded from below, and it is assumed that $e^{-N V(x)}$ exponentially damps as $|x|\to \infty$. 
Then, we prove that the expectation values concerning $a_0^2$ given in (\ref{pro_coupling})  
\be
\vev{a_0^{2k}} = \frac{\int d^{N^2}M \, a_0^{2k} \, e^{-S_{MM}}}{\int d^{N^2}M \, e^{-S_{MM}}}
\qquad (k=1, 2, \cdots)
\label{vevs_a}
\ee
vanish as $\epsilon \to 0$ in the case of finite $N$. 

After the angular integrations, integrals over the eigenvalues $\lambda_i$  ($i=1, \cdots, N$) remain in (\ref{vevs_a}): 
\bea
\vev{a_0^{2k}} & = & 
\frac{1}{Z_N}\int\left(\prod_{i=1}^N d\lambda_i\, e^{-NV(\lambda_i)}\right) \Delta(\lambda)^2 
\left(\frac{\epsilon^2}{(\frac{1}{N}\sum_i\lambda_i)^2+ \epsilon^2}\right)^k , \\
Z_N & = & \int\left(\prod_{i=1}^N d\lambda_i\, e^{-NV(\lambda_i)}\right) \Delta(\lambda)^2,  
\eea
where $\Delta(\lambda)$ is the Vandermonde determinant and can be written as 
\bea
\Delta(\lambda)^2 & = & \prod_{i>j} (\lambda_i-\lambda_j)^2 \le \prod_{i>j}(|\lambda_i|+|\lambda_j|)^2 \nono \\
 & = & \sum_{n_1,\cdots, n_N =0}^{2(N-1)}C_{n_1,\cdots,n_N} |\lambda_1|^{n_1}\cdots |\lambda_N|^{n_N}
 \eea
with $C_{n_1,\cdots,n_N}$ nonnegative finite numbers. 
Thus, 
\bea
\vev{a_0^{2k}} & \le &  \frac{1}{Z_N}\sum_{n_1,\cdots, n_N =0}^{2(N-1)}C_{n_1,\cdots,n_N} 
\int \left(\prod_{i=1}^N d\lambda_i\, e^{-NV(\lambda_i)}\right) \nono \\
 & & \hspace{2cm}\times |\lambda_1|^{n_1}\cdots |\lambda_N|^{n_N} 
\left(\frac{(N\epsilon)^2}{(\lambda_1+\cdots +\lambda_N)^2+ (N\epsilon)^2}\right)^k. 
\eea 

First, let us evaluate the integral of $\lambda_1$: 
\be
I_{n_1} \equiv \int d\lambda_1 \, e^{-NV(\lambda_1)} |\lambda_1|^{n_1}
\left(\frac{(N\epsilon)^2}{(\lambda_1+\cdots +\lambda_N)^2+ (N\epsilon)^2}\right)^k.
\ee
From the assumptions for the potential $V$, there exists a finite positive number depending on $n_1$ and $N$: $M(n_1, N)$ 
satisfying 
\be
e^{-NV(\lambda_1)}|\lambda_1|^{n_1} \le M(n_1, N) \qquad {\rm for} \qquad \forall \lambda_1. 
\ee
(Of course, $M(n_1, N)$ depends also on coupling constants of $V$. We suppressed such dependences here.)
Then, we obtain the bound
\bea
I_{n_1} & \le & M(n_1,N)\int_{-\infty}^{\infty}d\lambda_1 \, 
\left(\frac{(N\epsilon)^2}{(\lambda_1+\cdots +\lambda_N)^2+ (N\epsilon)^2}\right)^k \nono \\
 & = & M(n_1, N) \int_{-\infty}^{\infty}d\lambda \, 
\left(\frac{(N\epsilon)^2}{\lambda^2+ (N\epsilon)^2}\right)^k \nono \\
 & = & \pi N\epsilon \frac{(2k-3)!!}{(2k-2)!!} M(n_1, N). 
 \label{bound_In1}
 \eea

Using (\ref{bound_In1}), we have 
\bea
 0\le \vev{a_0^{2k}} & \le & \frac{1}{Z_N} \sum_{n_1,\cdots, n_N =0}^{2(N-1)}C_{n_1,\cdots,n_N} 
\int \left(\prod_{i=2}^N d\lambda_i\, e^{-NV(\lambda_i)}\right) I_{n_1}
|\lambda_2|^{n_2}\cdots |\lambda_N|^{n_N}  \nono \\
 & \le & 
\frac{1}{Z_N}\pi N\epsilon \, \frac{(2k-3)!!}{(2k-2)!!}\sum_{n_1,\cdots, n_N =0}^{2(N-1)}C_{n_1,\cdots,n_N} 
 M(n_1,N) J_{n_2}\cdots J_{n_N}, 
 \label{hyouka}
 \eea
 where 
 \be
 J_n \equiv \int d\lambda \, e^{-NV(\lambda)} |\lambda|^n. 
 \ee
For finite $N$, since $M(n_1, N)$, $J_n$ and $1/Z_N$ are finite and the sum over $n_1, \cdots, n_N$ is a finite sum, 
the r.h.s. of (\ref{hyouka}) becomes zero as ${\cal O}(\epsilon)$ when $\epsilon \to 0$. 
Hence, it is shown that    
\be
\vev{a_0^{2k}} \to 0 \qquad (\epsilon\to 0) \qquad \mbox{for $N$: finite}.  
\ee

\section{Perturbative Calculation in SUSY Matrix Model}
\label{app:loop_susymat}
\setcounter{equation}{0}
In this appendix, we explain the perturbative loop calculation in the supersymmetric matrix model 
sector leading to (\ref{inducedK2}) and (\ref{inducedW2}). 
Let us evaluate the integrals 
\bea 
e^{-\Gamma_{\rm ind}}  & = & \int (d{\cal M}\, d\bar{\cal M}) \, e^{-\tilde{S}}, \nono \\
\tilde{S} & \equiv & -S_{MM} -\Delta S_{MM} -S_{int}, 
\eea
up to the quadratic orders of $h, \bar{h}$. 
For notational simplicity, we put 
\be
J=\frac{f}{N}\int d^4x\, \Phi_0 = \phi_J + \sqrt{2}\theta\psi_J + \theta\theta F_J, 
\ee
and write $S_{int}$ as 
\be
S_{int} = \tr\left[\int d^2\theta \, J{\cal M} + \int d^2\bar{\theta} \, \bar{J} \bar{\cal M} \right]. 
\ee
Note that the contour of the integrals for the auxiliary variables 
$F_M, \bar{F}_M$ are taken along the imaginary axis. 
Actually, the action $-S_{MM}$ is expressed in terms of the component variables as  
\bea
-S_{MM} & = & \tr \left[-\int d^2\theta d^2\bar{\theta} \, \bar{\cal M}{\cal M} 
- \frac{m}{2} \int d^2\theta \, {\cal M}^2 - \frac{\bar{m}}{2} \int d^2\bar{\theta} \, \bar{\cal M}^2\right] \nono \\
 & = & \tr \left[-(\bar{F}_M+mM)(F_M+\bar{m}\bar{M}) +m\bar{m}M\bar{M} + \frac{m}{2}\chi\chi 
+ \frac{\bar{m}}{2}\bar{\chi}\bar{\chi} \right],  
\eea
which shows that the integrals along the above-mentioned contour are well-defined. 

Under the shift 
\bea
{\cal M}& \to &{\cal M} - \frac{1}{m} J+ \frac{1}{m\bar{m}}\bar{F}_J, \nono \\
\bar{\cal M}& \to &\bar{\cal M} - \frac{1}{\bar{m}} \bar{J}+ \frac{1}{m\bar{m}}F_J 
\eea
to cancel the linear terms with respect to ${\cal M}, \bar{\cal M}$ in ${\cal O}(h^0, \bar{h}^0)$ part of $\tilde{S}$, 
we have 
\bea
\lefteqn{S_{MM} + S_{int}  =  \tr \left[\int d^2\theta d^2\bar{\theta} \, \left(\bar{\cal M} {\cal M} + \frac{1}{m\bar{m}}\bar{J}J
                                \right) \right.} \nono \\
 & & \left. + \int d^2\theta\, \left(\frac{m}{2} {\cal M}^2 -\frac{1}{2m}J^2\right) 
        + \int d^2\bar{\theta}\, \left(\frac{\bar{m}}{2} \bar{\cal M}^2 -\frac{1}{2\bar{m}}\bar{J}^2\right)\right], 
\eea
\bea
\Delta S_{MM} & = & \tr \left[\frac{h}{3} \int d^2\theta\, \left\{ {\cal M}^3 
     -\frac{3}{m}{\cal M}^2A
     + \frac{3}{m^2} {\cal M}A^2 -\frac{1}{m^3}A^3 \right\} \right.       \nono \\
    & & \hspace{7mm}\left.  +\frac{\bar{h}}{3} \int d^2\bar{\theta}\, \left\{ \bar{\cal M}^3 
     -\frac{3}{\bar{m}}\bar{\cal M}^2\bar{A} 
     + \frac{3}{\bar{m}^2} \bar{\cal M}\bar{A}^2 -\frac{1}{\bar{m}^3}\bar{A}^3 \right\} \right], 
\eea
with 
\be
A\equiv J-\frac{1}{\bar{m}}\bar{F}_J, \qquad 
\bar{A} \equiv \bar{J}-\frac{1}{m}F_J. 
\ee
Then, the integrals can be written as 
\bea
e^{-\Gamma_{\rm ind}} & = & \exp\left[\tr \left\{\int d^2\theta d^2\bar{\theta}\, \frac{1}{m\bar{m}}J\bar{J} 
   + \int d^2\theta \, \left(-\frac{1}{2m}J^2 -\frac{h}{3m^3}A^3\right) \right.\right. \nono \\
    & & \hspace{2cm} \left.\left. +   \int d^2\bar{\theta} \, \left(-\frac{1}{2\bar{m}}\bar{J}^2 
             -\frac{\bar{h}}{3\bar{m}^3}\bar{A}^3\right) \right\} \right] \nono \\
    & & \times Z_0 \vev{e^{I + \bar{I}}}_0, 
\label{exp-Gamma}    
\eea
where 
\bea
 & & \vev{\cal O}_0  \equiv  \frac{1}{Z_0} \int  (d{\cal M} \, d\bar{\cal M}) \, e^{S_{MM}} {\cal O}, \qquad
Z_0  \equiv     \int  (d{\cal M} \, d\bar{\cal M}) \, e^{S_{MM}}, \\
 & & I  \equiv  \tr \left[ \int d^2\theta \, \left\{ \frac{h}{3} {\cal M}^3 
                     -\frac{h}{m} {\cal M}^2 A
                     +\frac{h}{m^2} {\cal M} A^2\right\}\right], \nono \\
 & & \bar{I} \equiv  \tr \left[ \int d^2\bar{\theta} \, \left\{ \frac{\bar{h}}{3} \bar{\cal M}^3 
                     -\frac{\bar{h}}{\bar{m}} \bar{\cal M}^2 \bar{A} 
                     +\frac{\bar{h}}{\bar{m}^2} \bar{\cal M} \bar{A}^2\right\}\right]. 
\eea
Since $Z_0$ in (\ref{exp-Gamma}) is an irrelevant constant independent of $m$ and $\bar{m}$, 
we will obtain the expression of $\Gamma_{\rm ind}$ by evaluating 
$\vev{e^{I + \bar{I}}}_0$ in the perturbative expansion of  $h$, $\bar{h}$. 

In order to get the propagators for supervariables ${\cal M}$, $\bar{\cal M}$, first let us compute the propagators for 
component variables: 
\bea
 & & \vev{M_{ij}\bar{M}_{kl}}_0 =  \frac{\int (dM d\bar{M})\, e^{-m\bar{m}\,\tr (M\bar{M})} M_{ij}\bar{M}_{kl}}{\int (dM d\bar{M})\, 
                                        e^{-m\bar{m}\,\tr (M\bar{M})}} = \frac{1}{m\bar{m}}\delta_{il}\delta_{jk}, \\           
 & & \vev{(\chi_\alpha)_{ij}(\chi_\beta)_{kl}}_0 =  \frac{\int (d\chi) e^{-\frac{m}{2}\tr (\chi\chi)} (\chi_\alpha)_{ij}(\chi_\beta)_{kl}}{\int
                                            (d\chi) e^{-\frac{m}{2}\tr (\chi\chi)}} =  -\frac{1}{m}\varepsilon_{\alpha\beta}\delta_{il}\delta_{jk}.
\eea
Similarly, 
\be
\vev{(\bar{\chi}_{\dot{\alpha}})_{ij}(\bar{\chi}_{\dot{\beta}})_{kl}}_0 = \frac{1}{\bar{m}} \varepsilon_{\dot{\alpha}\dot{\beta}}
\delta_{il}\delta_{jk}.
\ee          
Also, 
\bea
\vev{(F_M)_{ij}(\bar{F}_M)_{kl}}_0  
 & = &  \frac{\int (dM dF_M d\bar{M}d\bar{F}_M) \, 
e^{\tr [\bar{F}_MF_M+mMF_M+\bar{m}\bar{M}\bar{F}_M]} \,(F_M)_{ij}(\bar{F}_M)_{kl}}{\int 
(dM dF_M d\bar{M}d\bar{F}_M) \, e^{\tr [\bar{F}_MF_M+mMF_M+\bar{m}\bar{M}\bar{F}_M]}} \nono \\
 & =&  \frac{\int  (dM dF_M d\bar{M}d\bar{F}_M) \, e^{\tr \bar{F}_MF_M}e^{-m\bar{m}\,\tr (\bar{M}M)} \,(F_M-\bar{m}\bar{M})_{ij}
(\bar{F}_M-mM)_{kl}}{\int  (dM dF_M d\bar{M}d\bar{F}_M) \, e^{\tr \bar{F}_MF_M}e^{-m\bar{m}\,\tr (\bar{M}M)}} \nono \\
 & =&  \frac{\int (dM d\bar{M}) \, e^{-m\bar{m} \,\tr \bar{M}M}\,[-\delta_{il}\delta_{jk} + m\bar{m}\bar{M}_{ij}M_{kl}]}{\int 
(dM d\bar{M}) \, e^{-m\bar{m} \,\tr \bar{M}M}} \nono \\
 & =&  -\delta_{il}\delta_{jk} + m\bar{m}\frac{1}{m\bar{m}}\delta_{il}\delta_{jk} = 0, \\
\vev{M_{ij}(F_M)_{kl}}_0  
 & = &  \frac{\int (dM dF_M d\bar{M}d\bar{F}_M) \, 
e^{\tr [\bar{F}_MF_M+mMF_M+\bar{m}\bar{M}\bar{F}_M]} \,M_{ij}(F_M)_{kl}}{\int (dM dF_M d\bar{M}d\bar{F}_M) \, 
e^{\tr [\bar{F}_MF_M+mMF_M+\bar{m}\bar{M}\bar{F}_M]}} \nono \\
 & =&   \frac{\int  (dM dF_M d\bar{M}d\bar{F}_M) \, e^{\tr \bar{F}_MF_M}e^{-m\bar{m}\,\tr (\bar{M}M)} \,M_{ij}
(F_M-\bar{m}\bar{M})_{kl}}{\int  (dM dF_M d\bar{M}d\bar{F}_M) \, e^{\tr \bar{F}_MF_M}e^{-m\bar{m}\,\tr (\bar{M}M)}} \nono \\
 &= &   \frac{\int (dM d\bar{M}) \, e^{-m\bar{m} \,\tr \bar{M}M}\,(-\bar{m})M_{ij}\bar{M}_{kl}}{\int 
(dM d\bar{M}) \, e^{-m\bar{m} \,\tr \bar{M}M}} \nono \\
 &= &   (-\bar{m})\frac{1}{m\bar{m}}\delta_{il}\delta_{jk}  =  \frac{-1}{m}\delta_{il}\delta_{jk}, \\
\vev{\bar{M}_{ij}(\bar{F}_M)_{kl}}_0 & =  & \frac{-1}{\bar{m}}\delta_{il}\delta_{jk}.
\eea
The other propagators vanish. 
 
The above results are combined into the form of the supervariable propagators as 
\bea
\vev{{\cal M}(\theta)_{ij} \bar{\cal M}(\bar{\theta}')_{kl}}_0 & = &  \vev{M_{ij}\bar{M}_{kl}}_0 = \frac{1}{m\bar{m}}\delta_{il}\delta_{jk}, 
\label{calM-pro1}\\
\vev{{\cal M}(\theta)_{ij} {\cal M}(\theta')_{kl}}_0 & = & 2\vev{(\theta\chi_{ij})(\theta'\chi_{kl})}_0 
  + \theta\theta\vev{(F_M)_{ij}M_{kl}}_0 + \theta'\theta'\vev{M_{ij}(F_M)_{kl}}_0 \nono \\
   & = & -2\theta^\alpha\theta'^\beta\frac{-1}{m}\varepsilon_{\alpha\beta}\delta_{il}\delta_{jk} 
           + (\theta\theta+\theta'\theta')\frac{-1}{m}\delta_{il}\delta_{jk} \nono \\
   & = & \delta(\theta-\theta')\frac{-1}{m}\delta_{il}\delta_{jk}, 
\label{calM-pro2}   
\eea
where the notation ${\cal M}(\theta)$, $\bar{\cal M}(\bar{\theta})$ is used in order to manifest 
the $\theta$, $\bar{\theta}$-dependence of ${\cal M}, \bar{\cal M}$. 
Here $\delta(\theta) \equiv \theta\theta$, in particular $\delta(0)=0$. 
Similarly, 
\be
\vev{\bar{\cal M}(\bar{\theta})_{ij} \bar{\cal M}(\bar{\theta}')_{kl}}_0 = 
                 \delta(\bar{\theta}-\bar{\theta}')\frac{-1}{\bar{m}}\delta_{il}\delta_{jk}. 
\label{calM-pro3}                 
\ee

We compute $\vev{e^{I+\bar{I}}}_0$ by the Wick contractions 
using the supervariable propagators (\ref{calM-pro1}), (\ref{calM-pro2}), and (\ref{calM-pro3}). 
First, from (\ref{exp-Gamma}), 
\be
\left({\cal O}(h^0, \bar{h}^0\right) \mbox{ of } -\Gamma_{\rm ind}) 
= \tr \left[\int d^2\theta d^2\bar{\theta}\, \frac{1}{m\bar{m}}J\bar{J}
-\frac{1}{2m}\int d^2\theta \, J^2 -\frac{1}{2\bar{m}}\int d^2\bar{\theta} \, \bar{J}^2 \right].
\label{Gamma_h0}
\ee
Note that $\vev{\tr \int d^2\theta \, {\cal M}(\theta)^2 A(\theta)}_0=0$, 
because of  $\delta(0)$ appearing from the contraction of ${\cal M}$'s with the same $\theta$. 
Thus, 
\be
\left({\cal O}(h, \bar{h}\right) \mbox{ of } \ln \vev{e^{I+\bar{I}}}_0) = 0, 
\ee
and 
\be
\left({\cal O}(h, \bar{h}) \mbox{ of } -\Gamma_{\rm ind}\right) = \tr\left[-\frac{h}{3m^3}\int d^2\theta A^3
       -\frac{\bar{h}}{3\bar{m}^3}\int d^2\bar{\theta} \, \bar{A}^3\right]. 
\label{Gamma_h1}       
\ee

Next, we calculate 
\be
\left({\cal O}(h^2) \mbox{ of }-\Gamma_{\rm ind} \right) = \frac12 \vev{I^2}_{0, C},
\ee
where the suffix $C$ means taking contributions from the connected graphs. 
It can be seen that 
\be
\int d^2\theta \int d^2\theta' \, \vev{\tr \left[{\cal M}(\theta)^nA(\theta)\right]\tr \left[{\cal M}(\theta')^nA(\theta')\right]}_{0,C} 
= 0 \qquad \mbox{for }n\ge 2
\ee
due to appearance of $\delta(0)$, and the $Z_2$-symmetry of $S_{MM}$ under  
\be
{\cal M}\to -{\cal M}, \qquad \bar{\cal M}\to -\bar{\cal M}
\ee
implies 
\be
\vev{{\cal M}(\theta_1)_{i_1j_1} \cdots {\cal M}(\theta_n)_{i_nj_n} 
\bar{\cal M}(\bar{\theta}_1)_{k_1l_1} \cdots \bar{\cal M}(\bar{\theta}_m)_{k_ml_m}}_0 =0
\qquad \mbox{for $n+ m$: odd}. 
\ee
{}From these identities, we find 
\bea
\left({\cal O}(h^2) \mbox{ of }-\Gamma_{\rm ind} \right) & = & \frac12 \frac{h^2}{m^4} \int d^2\theta \int d^2\theta' \,
\vev{\tr \left[{\cal M}(\theta)A(\theta)^2\right]\tr \left[{\cal M}(\theta')A(\theta')^2\right]}_{0,C} \nono \\
 & = & -\frac{Nh^2}{2m^5} \int d^2\theta \, A^4
\label{Gamma_h2} 
\eea
and 
\be
\left({\cal O}(\bar{h}^2) \mbox{ of }-\Gamma_{\rm ind} \right)  =  -\frac{N\bar{h}^2}{2\bar{m}^5} \int d^2\bar{\theta} \, \bar{A}^4. 
\label{Gamma_hb2}
\ee
Also, due to 
\bea
 & & \int d^2\theta\int d^2\bar{\theta}\, \vev{\tr \left[{\cal M}(\theta)A(\theta)^2\right] 
\tr \left[\bar{\cal M}(\bar{\theta})\bar{A}(\bar{\theta})^2\right]}_{0, C} 
 = \int d^2\theta\int d^2\bar{\theta}\, \frac{N}{m\bar{m}}\,A^2\bar{A}^2, \nono \\
 & & \int d^2\theta\int d^2\bar{\theta}\, \vev{\tr \left[{\cal M}^2(\theta)A(\theta)\right] 
\tr \left[\bar{\cal M}^2(\bar{\theta})\bar{A}(\bar{\theta})\right]}_{0, C}  
=  \int d^2\theta\int d^2\bar{\theta}\, \frac{2N^2}{m^2\bar{m}^2}\, A\bar{A}, \nono \\
 & & \int d^2\theta\int d^2\bar{\theta}\, \vev{\tr {\cal M}(\theta)^3 \tr \bar{\cal M}(\bar{\theta})^3}_{0, C} 
=\int d^2\theta\int d^2\bar{\theta} \, (\mbox{$\theta$, $\bar{\theta}$-independent terms}) =  0, \nono \\
& & 
\eea
we obtain 
\be
\left({\cal O}(h\bar{h}) \mbox{ of }-\Gamma_{\rm ind} \right)= \vev{I\bar{I}}_{0,C}
 = \frac{h\bar{h}}{m^3\bar{m}^3}\int d^2\theta\int d^2\bar{\theta}\,
  \left[NA^2\bar{A}^2+2N^2A \bar{A} \right]. 
\label{Gamma_hhb}  
\ee

Gathering the results (\ref{Gamma_h0}), (\ref{Gamma_h1}), (\ref{Gamma_h2}), (\ref{Gamma_hb2}), and (\ref{Gamma_hhb}), 
we have the expression of $\Gamma_{\rm ind}$, which can be written in terms of $A, \bar{A}$ by  
\bea
\Gamma_{\rm ind} & = & \int d^2\theta d^2\bar{\theta} \left\{\left(\frac{N}{m\bar{m}}
-2\frac{h\bar{h}N^2}{m^3\bar{m}^3}\right)A\bar{A} - \frac{h\bar{h}N}{m^3\bar{m}^3}A^2\bar{A}^2\right\} \nono \\
& &
+\int d^2\theta \left(\frac{N}{2m}A^2 + \frac{hN}{3m^3}A^3 + \frac{h^2N}{2m^5}A^4\right) \nono \\
& & 
+\int d^2\bar{\theta}\left(\frac{N}{2\bar{m}}\bar{A}^2 + \frac{\bar{h}N}{3\bar{m}^3}\bar{A}^3 
+ \frac{\bar{h}^2N}{2\bar{m}^5}\bar{A}^4\right) \nono \\
 & & +  {\cal O}(h^3, \bar{h}^3, h^2\bar{h}, h\bar{h}^2). 
\eea
Thus, we see that the potential part of the total action $\Gamma = -S_{OR} + \Gamma_{\rm ind}$, given by 
\be
\Gamma^{(0)} = V_{4D} \left(-\int d^2\theta d^2\bar{\theta} \, {\cal K} -\int d^2\theta \, {\cal W} 
-\int d^2\bar{\theta} \, \bar{\cal W} \right), 
\ee
is expressed as in (\ref{inducedK2}) and (\ref{inducedW2}), using the relation 
\be
A = \frac{f}{N}V_{4D} \tilde{\Phi}_0^{(0)}, \qquad 
\bar{A} = \frac{\bar{f}}{N}V_{4D} \bar{\tilde{\Phi}}_0^{(0)}. 
\ee 

\paragraph{Remarks on reduced matrix models on the Minkowski space} 
We explain that the matrix model can not consistently couple to field theory sector 
in the Minkowski space. 
In the Minkowski case, the induced action is defined by 
\be
e^{i\Gamma_{\rm ind}^{(M)}} \equiv \int (d{\cal M} \, d\bar{\cal M}) \, e^{iS_{MM} + i\Delta S_{MM} + iS_{int}}, 
\ee
and the propagators are 
\bea
\vev{{\cal M}(\theta)_{ij}\bar{\cal M}(\bar{\theta}')_{kl}}_0 & = & \frac{-i}{m\bar{m}}\delta_{il}\delta_{jk}, \nono \\
\vev{{\cal M}(\theta)_{ij} {\cal M}(\theta')_{kl}}_0 & = & \delta(\theta-\theta') \frac{i}{m}\delta_{il}\delta_{jk}, \nono \\
\vev{\bar{\cal M}(\bar{\theta})_{ij} \bar{\cal M}(\bar{\theta}')_{kl}}_0 & = & 
\delta(\bar{\theta}-\bar{\theta}') \frac{i}{\bar{m}}\delta_{il}\delta_{jk},
\eea
instead of (\ref{calM-pro1}), (\ref{calM-pro2}), and (\ref{calM-pro3}). 
The calculation at the tree level is not problematic, but at the one-loop order we have 
\be
\int d^2\theta\int d^2\bar{\theta}\, \vev{\tr \left[{\cal M}^2(\theta)A(\theta)\right] 
\tr \left[\bar{\cal M}^2(\bar{\theta})\bar{A}(\bar{\theta})\right]}_{0, C} 
=  \int d^2\theta\int d^2\bar{\theta}\, \frac{-2N^2}{m^2\bar{m}^2}\, A \bar{A}, 
\label{minkow_oneloop} 
\ee
and hence 
\be
\left({\cal O}(h\bar{h}) \mbox{ of }i\Gamma_{\rm ind}^{(M)}\right)
=   i\frac{h\bar{h}}{m^3\bar{m}^3}\int d^2\theta\int d^2\bar{\theta}\,
  \left[N A^2\bar{A}^2-i2N^2 A \bar{A} \right]. 
\label{hhb_min}
\ee
The first term of (\ref{hhb_min}), which is from the tree diagram, gives a real-valued contribution to $\Gamma_{\rm ind}^{(M)}$, 
whereas the second term from the one-loop (\ref{minkow_oneloop}) yield pathological imaginary-valued contributions. 
It is due to the fact that the matrix model has no momentum modes. 
For more detailed explanation, let us consider, say, 
a $D$-dimensional matrix field theory with the propagator $\frac{i}{p^2-m^2}$ and the ($n$-point) vertex $-ig$. 
The contribution from a $L$-loop diagram with $I$ internal lines and $V$ vertices has the form: 
\bea
\lefteqn{G_L \equiv 
(-ig)^V \int \frac{d^Dk_1}{(2\pi)^D}\cdots \frac{d^Dk_L}{(2\pi)^D} \, \frac{i}{P_1^2-m^2}\cdots \frac{i}{P_I^2-m^2}} \nono \\
 & &= i^{L-1}g^V  \int \frac{d^Dk_1}{(2\pi)^D}\cdots \frac{d^Dk_L}{(2\pi)^D} \, \frac{1}{P_1^2-m^2}\cdots \frac{1}{P_I^2-m^2}, 
\eea 
where the internal momenta $P_i$ ($i=1, \cdots, I$) are linear functions of the loop momenta $k_\ell$ ($\ell=1, \cdots, L$), 
and the relation $L=I-V+1$ was used. 
For simplicity, we omit the $N$-dependence which is irrelevant in this argument. 
The integrals are assumed to be properly cutoff in the ultraviolet region to regulate possible divergences. 
At first sight, the integrations of the loop momenta in the r.h.s. might give a real value and the value of the diagram might be 
\be
G_L \stackrel{?}{=} i^{L-1}\times (\mbox{real quantity}) . 
\label{GL?}
\ee
However, this is not true. In order to evaluate the integrals in a well-defined way, 
we need to take the so-called $i\varepsilon$-prescription, or equivalently 
to do the Wick rotation of the time-component of the momenta. In fact, after the Wick rotation, we have 
\bea
G_L  & = & i(-g)^V \int \frac{d^Dk_1}{(2\pi)^D}\cdots \frac{d^Dk_L}{(2\pi)^D} \, \frac{1}{P_1^2+m^2}\cdots \frac{1}{P_I^2+m^2} 
\nono \\
 & = & i\times  (\mbox{real quantity}),
\label{GLeuclid} 
\eea
which gives a correct contribution to the loop amplitudes. 
Thus, in the Minkowski space, reality of the loop integrals is not manifest, 
and the $i\varepsilon$-prescription or the Wick rotation for the momenta 
is necessary to obtain the correct evaluation of the loop amplitudes. 
Next, consider the case of the matrix model, which is a dimensional reduction of the above field theory in the Minkowski space. 
Since there are no momentum modes in the model, we 
will have an evaluation of the corresponding diagram like 
\be
G_L^{\rm (mat)} \equiv (-ig)^V \left(\frac{-i}{m^2}\right)^I =  i^{L-1}g^V  \left(\frac{-1}{m^2}\right)^I. 
\ee
Here, the diagram has the structure $i^{L-1}\times (\mbox{real quantity})$, that agrees with the one-loop result (\ref{minkow_oneloop}). 
It gives imaginary-valued contributions to the induced action from the diagrams with the odd number of loops, and thus it 
violates the unitarity not to yield a consistent coupling between the matrix model and the field theory. 

On the other hand, in the case of the Euclid space, 
both of the field theory and its reduced matrix model lead to diagrams explicitly real-valued, 
and then the coupling between the field theory sector and the matrix model sector gives real-valued contributions 
to the induced action, which does not cause the above problem.

\section{Computation of the Vacuum Decay Rate}
\label{app:lifetime}
\setcounter{equation}{0}
In general, the decay rate of a false vacuum to the true vacuum is evaluated from the bounce solutions of 
the Euclidean action. 
Since the potential (\ref{eff_pot}) is too complicated to obtain the corresponding bounce solutions exactly, 
we use the method of ref.~\cite{Duncan:1992ai} in which the potential shape is approximated to a triangular or square shape. 
Because our potential is slowly varying for small $|\epsilon|$, the approximation is expected to give a reasonable order estimate. 
For simplicity, we consider the case of the parameters $m, \bar{m}, f, \bar{f}, \lambda, \bar{\lambda}$ being real, so that the bounce solutions will be real. 
 
First, in order to approximate the potential (\ref{eff_pot}) to 
a triangular shape, 
we must calculate the position and height of a peak. 
Before doing it, we rescale as 
\be
\phi_X \to \frac{1}{g'}\phi_X, \qquad \lambda' \to \frac{1}{g'}\lambda', 
\ee
so that $|g'|^2$ plays the role of $\hbar$, which makes it easy 
to see the loop effects.  
Then, the tree level potential becomes 
\be
V_{\rm tree} = \frac{1}{|g'|^2}|\lambda'-\epsilon \phi_X|^2,
\ee
and $V_{\rm CW}={\cal O}(|g'|^0)$ with the variables $x, z$ in (\ref{variables_xz}) becoming 
\be
x=2\left|\frac{\phi_X}{\mu}\right|^2, \qquad z=\left|(\lambda'-\epsilon \phi_X)\frac{1}{\mu^2}\right|. 
\ee
The local minimum is expressed as 
\be
\phi_{X \, {\rm min}} = \frac{R}{2\tilde{m}_X^2}\frac{\bar{\epsilon}}{\bar{\lambda}'} 
\ee
with the height 
\be
V_0 = |\lambda'|^2\left[\frac{1}{|g'|^2}+ \frac{1}{8\pi^2}\left(\log \frac{|\mu|^2}{M_{\rm cutoff}^2} + \frac32 + v(y)\right)\right],
\ee
where the constants are 
\bea
 & & y = 2\left|\frac{\lambda'}{\mu^2}\right|, \qquad 
\tilde{m}_X^2 =  \frac{1}{2\pi^2}\left|\frac{\lambda'^2}{\mu^2}\right|\nu(y), \nono \\
 & & R =  2V_0 + \frac{|\lambda'|^2}{8\pi^2}\left(1-\frac{1+y}{y^2}\log (1+y)-\frac{1-y}{y^2}\log (1-y)\right). 
\eea
We consider the case $y={\cal O}(1)$ and 
\be
\left|\frac{\epsilon}{\mu}\right| \ll |g'^2| \ll 1
\label{epsilon_g'}
\ee
in which $\phi_{X \, {\rm min}}$ is near the origin, namely 
the dimensionless combination $|\frac{\phi_{X \, {\rm min}}}{\mu}|$ is sufficiently smaller than one\footnote{
Note that ${\cal O}\left(\left|\mu^2\right|\right) = {\cal O}\left(\left|\lambda'\right|\right)$ because of $y={\cal O}(1)$.}.  

{}From the fact that the Coleman-Weinberg potential of the original O'Raifeartaigh model, which is obtained by setting 
$\epsilon =0$ in (\ref{explicit_CW}), is a monotonically increasing function of $x$~\cite{Intriligator:2007cp}, 
it is seen that the point 
$\phi_{X \, {\rm peak}}$ giving the peak of $V_{\rm eff}^{(1)}$ moves to the infinity as $\epsilon \to 0$. 
Next, let us look at the expansion of the potential $V_{\rm eff}^{(1)}= V_{\rm tree} + V_{\rm CW}$ around $\epsilon =0$: 
\bea
V_{\rm tree} & = & \left|\frac{\lambda'}{g'}\right|^2 -\frac{2}{|g'|^2} \, \mbox{Re}\left(\bar{\lambda}'\epsilon \phi_X\right) 
+\frac{1}{|g'|^2}|\epsilon\phi_X|^2, \label{Vtree_exp}\\
V_{\rm CW} & = & V_{\rm CW} |_{\epsilon=0} + {\cal O}\left(\left|\epsilon\phi_X\mu^2\right|\right). 
\label{VCW_exp}
\eea
Note that $\epsilon$ always appears through the combination 
$\epsilon\phi_X$. 
In order to change the monotonically increasing behavior of $V_{\rm CW} |_{\epsilon=0}$ to make a peak in $V_{\rm eff}^{(1)}$, 
the second term in r.h.s. of (\ref{Vtree_exp}) must become comparable to $V_{\rm CW} |_{\epsilon=0}$.  
Therefore, it is plausible to assume 
\be
\phi_{X \, {\rm peak}}\sim {\cal O}\left(\left|\frac{g'^2}{\epsilon}\mu^2\right|\right). 
\label{assume_X}
\ee
When we find $\phi_{X \, {\rm peak}}$ by solving the equation 
\be
\frac{\partial}{\partial\phi_X}\left(V_{\rm tree} + V_{\rm CW}\right)=0,
\label{eq_peak}
\ee
we can neglect the $|\phi_X|^2$ term in $V_{\rm tree}$ because 
\be
\frac{1}{|g'|^2}|\epsilon\phi_{X\, {\rm peak}}|^2 = {\cal O}\left(|g'^2\mu^4|\right)
\ee
is the same order as the two-loop contribution which is already neglected by the condition (\ref{epsilon_g'}). 
Likewise, we do not have to consider the $\epsilon$-dependent terms in $V_{\rm CW}$, since 
${\cal O}(|\epsilon\phi_{X\, {\rm peak}}\, \mu^2|)$ is equal to the order of the two-loop contribution again. 
Furthermore, noting that (\ref{assume_X}) is large because of (\ref{epsilon_g'}), it is reasonable 
to replace $V_{\rm CW}|_{\epsilon=0}$ with its asymptotic form 
\be
V_{\rm CW}|_{\epsilon=0} \sim \frac{|\lambda'|^2}{8\pi^2}\left(\log \frac{|2\phi_X|^2}{M_{\rm cutoff}^2}+ \frac32\right) 
\ee 
for $|\phi_X|$ large. 
Making use of all the above simplifications reduces the equation (\ref{eq_peak}) to 
\be
-\frac{\bar{\lambda}'\epsilon+\lambda'\bar{\epsilon}}{|g'|^2} + \frac{|\lambda'|^2}{4\pi^2}\frac{1}{\phi_X}=0. 
\ee  
Hence, we have 
\be
\phi_{X \, {\rm peak}}= \frac{\lambda'}{8\pi^2}\frac{|g'|^2}{\epsilon}. 
\label{X_peak_sol}
\ee
The solution (\ref{X_peak_sol}) is consistent with the assumption (\ref{assume_X}). 
Also, in order for (\ref{X_peak_sol}) to satisfy the condition 
$z<\frac12$ which corresponds to (\ref{nontachyonic}), 
it is sufficient to impose $y<1$.  
Then, the potential height of the peak is given by 
\be
V_{\rm peak} = V_{\rm eff}^{(1)}(\phi_{X \, {\rm peak}}) = 
V_0 + \frac{|\lambda'|^2}{4\pi^2}\log\left|\frac{\lambda'g'^2}{\epsilon \mu}\right| + {\cal O}(1). 
\ee

Approximating the potential curve to the triangular shape connecting the points 
$(\phi_{X \, {\rm min}}, V_0)$, $(\phi_{X \, {\rm peak}}, V_{\rm peak})$ and $(\frac{\lambda'}{\epsilon}, 0)$ 
as in ref.~\cite{Duncan:1992ai}, 
we consider the spherically symmetric bounce solution $\phi(r)$ satisfying 
\be
\ddot{\phi}(r) +\frac{3}{r}\dot{\phi}(r) = V_{\rm tri}'(\phi(r))
\label{bounce_sol}
\ee
with the boundary conditions 
\be
\lim_{r \to \infty} \phi(r) = \phi_{X \, {\rm min}}, 
\qquad \dot{\phi}(r=0) = 0. 
\ee
Here, $r = \sqrt{x_{\mu} x_{\mu}}$, the dot denotes the derivative with respect to $r$, 
$V_{\rm tri}$ represents the triangular potential obtained from $V_{\rm eff}^{(1)}$ and the prime 
means the differentiation with respect to $\phi$.   
Once such solution is found, the tunneling probability is given by 
\be
e^{-B}=\exp\left\{-S_E[\phi(r)]+S_E[\phi_{X \, {\rm min}}]\right\}
\ee
with 
\be
S_E[\phi] = \int d^4x\, \left[\frac12 (\partial_{\mu}\phi)^2 + V_{\rm tri}(\phi)\right]. 
\ee 
This is the contribution from the one-bounce configuration. Taking into account to multi-bounce configurations 
by using the dilute-gas approximation, we have $e^{-B}$ for the decay rate of the false vacuum. 
Hence, the lifetime is obtained as $e^B$. 

In ref.~\cite{Duncan:1992ai}, the two cases are discussed for the solutions of the triangular potential. 
Correspondingly, in our setting, we have 
\bea
B & \sim & \frac{8\pi^8}{3}\left|\frac{\lambda'^2}{\epsilon^4}\right|\frac{1}{\log\left|\frac{\lambda'g'^2}{\epsilon\mu}\right|} 
\qquad \mbox{for}\quad \frac{V_0}{V_{\rm peak}}\gtrsim \frac34, \nono \\
B & \sim & \frac{2}{3\pi^2}\frac{|\lambda'|^8}{|\epsilon|^4 V_0^3}\left(\log \left|\frac{\lambda'g'^2}{\epsilon\mu}\right|\right)^2
\qquad \mbox{for}\quad \frac{V_0}{V_{\rm peak}}\lesssim \frac34. 
\eea
For both cases, $B$ is proportional to $|\epsilon|^{-4}$ up to the logarithmic corrections. 
Thus, the lifetime is estimated as 
\be
e^B = {\cal O}\left(\exp\left\{|\epsilon|^{-4}\right\}\right)= {\cal O}\left(\exp\left\{\left(\frac{N}{V_{4D}}\right)^4\right\}\right), 
\ee
which becomes rapidly longer as $N$ increases. 

\paragraph{Remarks on bounce under nonlocal potential}
We estimated the lifetime of the metastable vacuum from the bounce solutions, which have a dependence on $r$. 
However, the effective potential up to the one-loop order $V_{\rm eff}^{(1)}$ is a function of the constant mode $\phi_X$, 
which is proportional to $\frac{1}{V_{4D}}\int d^4x\, \phi_0(x)$. 
It is a {\em nonlocal} potential of $\phi_0(x)$, and   
it is not completely clear the validity of the above computation starting with the triangular approximation of the {\em local} potential 
$V_{\rm eff}^{(1)}(\phi_X(x))$, which is a function of a local field $\phi_X(x)$: 
\be
\phi_X(x) = \phi_X + \mbox{(nonconstant modes)}.
\ee 
Here, we will give an argument guaranteeing the validity.  

Using the Taylor expansion 
\be
\phi_X(x') = \sum_{x_1, \cdots, x_4=0}^\infty \frac{1}{n_1!\cdots n_4!}(x'_1-x_1)^{n_1}\cdots (x'_4-x_4)^{n_4} 
\partial_{x_1}^{n_1}\cdots \partial_{x_4}^{n_4} \phi_X(x),
\ee
the constant mode $\phi_X = \frac{1}{V_{4D}}\int d^4x' \, \phi_X(x')$ can be expressed as 
\be
\phi_X = \phi_X(x) + \frac{1}{3!}L^2\partial_\mu\partial_\mu\phi_X(x) + {\cal O}(L^4\partial^4\phi_X(x)), 
\label{local_exp}
\ee
where we introduced the infra-red cutoff $L$ by $x_\mu \in [-L, L]$ and imposed the periodic boundary condition on 
$\phi_X(x)$ for each direction. 
Then, the nonlocal potential $V_{\rm eff}^{(1)}(\phi_X)$ is expanded by local potentials as 
\be
V_{\rm eff}^{(1)}(\phi_X) = V_{\rm eff}^{(1)}(\phi_X(x)) + \frac{1}{3!} V_{\rm eff}^{(1)'}(\phi_X(x)) L^2 \partial_\mu\partial_\mu \phi_X(x) 
+ {\cal O}(L^4\partial^4\phi_X(x)). 
\ee
We divide the Euclidean action 
$S_E[\phi_X(x)]=\int d^4x\,\left[\frac12(\partial_\mu\phi_X(x))^2 + V_{\rm eff}^{(1)}(\phi_X)\right]$ 
into the two parts: 
\bea
S_E[\phi_X(x)] & = & S_{loc}[\phi_X(x)] + S_{n.l.}[\phi_X(x)], \nono \\
S_{loc}[\phi_X(x)] & \equiv &  \int d^4x \, \left[\frac12(\partial_\mu\phi_X(x))^2 + V_{\rm eff}^{(1)}(\phi_X(x))\right] , \nono \\
S_{n.l.}[\phi_X(x)] & \equiv &  \int d^4x \, \left[\frac{1}{3!}V_{\rm eff}^{(1)'}(\phi_X(x))L^2\partial_\mu\partial_\mu\phi_X(x) 
+ {\cal O}(L^4\partial^4\phi_X(x))\right]. 
\eea  
Let us write a bounce solution of the equation of motion derived from $S_{loc}[\phi_X(x)]$ as $\phi_{\rm bounce}(r)$. 
In our case, by using the triangular approximation, 
\be
S_{loc}[\phi_{\rm bounce}(r)]-S_{loc}[\phi_{X \, {\rm min}}]={\cal O}\left(|\epsilon|^{-4}\right)
\label{local_bounce}
\ee
up to the logarithmic corrections. 
If the contribution $S_{n.l.}[\phi_{\rm bounce}(r)]$ is sufficiently small compared to (\ref{local_bounce}), 
it can be consistently neglected and the above computation of the vacuum decay rate is justified.  
We use the equation of motion and assume that the slope of the potential is not very large compared to that of the 
triangular potential to obtain 
\bea
S_{n.l.}[\phi_{\rm bounce}(r)]  & = &  \int d^4x\, \left[\frac{1}{3!} L^2 \left(V_{\rm eff}^{(1)'}(\phi_{\rm bounce}(r))\right)^2 
+ {\cal O}\left(L^4\partial^4\phi_{\rm bounce}(r)\right)\right] \nono \\
 & \lesssim & {\cal O}\left(V_{4D}^{3/2}|\epsilon|^2\right). 
\label{hyouka_S_nl}
\eea
Finally, $V_{4D} = (2L)^4$ was used. 

Comparing (\ref{local_bounce}) and (\ref{hyouka_S_nl}), for the justification, $|\epsilon|^{-1} \gg V_{4D}^{1/4}$ should be met. 
Hence, a slightly stronger condition for $N$ 
\be
N\gg V_{4D}^{5/4}
\label{N_jouken}
\ee
rather than $N\gg V_{4D}$ is necessary, 
and then the result of the lifetime is trusted when $N$ is large 
enough to satisfy (\ref{N_jouken}).


\end{document}